\documentclass[twoside,twocolumn,british,english,preprintnumbers,nofootinbib]{revtex4}
\usepackage{amsmath}
\usepackage{amssymb}
\usepackage{graphicx}
\usepackage{epsfig,bbm}
\usepackage{babel}
\usepackage{color}
\usepackage{hyperref}
\usepackage[normalem]{ulem}

\def\be{\begin{equation}}
\def\ee{\end{equation}}
\def\bea{\begin{eqnarray}}
\def\eea{\end{eqnarray}}

\def\mpl{M_{\rm P}}

\definecolor{amethyst}{rgb}{0.45, 0.3, 0.6}
\definecolor{darkgreen}{rgb}{0.0, 0.4, 0.26}

\begin{document}

	\title{Harmonic Hybrid Inflation}
	
	\author{Federico Carta}
	\email{federico.carta@desy.de}
	
	\author{Nicole Righi}
	\email{nicole.righi@desy.de}
	
	\author{Yvette Welling}
	\email{yvette.welling@desy.de}
	
	\author{Alexander Westphal}
	\email{alexander.westphal@desy.de}

	\affiliation{Deutsches Elektronen-Synchrotron DESY, Theory Group, D-22603 Hamburg, Germany}

	\begin{abstract}
		We present a mechanism for realizing hybrid inflation using two axion fields with a purely non-perturbatively generated scalar potential. The structure of the scalar potential is highly constrained by the discrete shift symmetries of the axions. We show that harmonic hybrid inflation generates observationally viable slow-roll inflation for a wide range of initial conditions. This is possible while accommodating certain UV arguments favoring constraints $f\lesssim \mpl$ and $\Delta\phi_{60}\lesssim \mpl$ on the axion periodicity and slow-roll field range, respectively. We discuss controlled $\mathbb{Z}_2$-symmetry breaking of the adjacent axion vacua as a means of avoiding cosmological domain wall problems. Including a minimal form of $\mathbb{Z}_2$-symmetry breaking into the minimally tuned setup leads to a prediction of primordial tensor modes with the tensor-to-scalar ratio in the range $10^{-4}\lesssim r \lesssim 0.01$, directly accessible to upcoming CMB observations. Finally, we outline several avenues towards realizing harmonic hybrid inflation in type IIB string theory.
	\end{abstract}

	\preprint{DESY-20-117}
	
	\maketitle
	
	\section{Introduction}	
		Inflation has become the leading paradigm to explain homogeneity and isotropy of our universe. The observations by the WMAP~\cite{Hinshaw:2012aka} and Planck~\cite{Akrami:2018odb} satellites, as well as the BICEP/Keck telescopes at the south pole~\cite{Ade:2018gkx}, provide us with powerful tools to explore the early stages of the cosmological history and to understand the inflationary epoch. Moreover, beyond serving as tests for many inflationary models proposed throughout the years, these data are also beginning to constrain top-down constructions attempting to embed inflation into a theory of quantum gravity. In fact, inflation is also one of the most promising means to explore the physics at energy scales requiring a candidate theory of quantum gravity such as string theory.
			
		Hybrid inflation, introduced by Linde in~\cite{Linde:1993cn}, is a mechanism of slow-roll inflaton which achieves the end of the slow-roll phase, driven by one scalar field, through an instability induced by the coupling with another scalar, which then undergoes a rapid `waterfall' roll to the minimum. The slow-roll phase itself is dominated by a large field-independent vacuum energy -- hence hybrid inflation `hybridizes' between `new' slow-roll inflation and `old' false-vacuum inflation. Interestingly, this vacuum energy domination implies that hybrid inflation possesses a small-ish field displacement corresponding to the last about 60 e-folds of observable inflation, $ \Delta\phi_{60} \lesssim\mpl $, which nevertheless does not become parametrically small. Hence, hybrid inflation naturally constitutes a mechanism realizing high-scale inflation which can accommodate the `Swampland Distance Conjecture' (SDC)~\cite{Ooguri:2006in}.
			
			Motivated by these features, we are proposing a regime of harmonic hybrid inflation driven by two axions acquiring a purely non-perturbative periodic scalar potential, and outline several roads towards realizing this regime in type IIB string theory.~\footnote{Such a regime may appear also in the many-axion setting studied in the `Axidental Universe' of~\cite{Bachlechner:2019vcb}.} Axions are an ubiquitous presence in most models of string compactifications, see e.g.~\cite{Svrcek:2006yi,Grimm:2007hs,Arvanitaki:2009fg,Carta:2020ohw}. String theory axions appear with an exponentially wide spectrum of masses suggesting some of them as suitable inflaton candidates. For inflationary models with axions see e.g.~\cite{Pajer:2013fsa,Baumann:2014nda} and references therein. We also note the very recent model of `hybrid monodromy inflation'~\cite{Kaloper:2020jso} which rather complementarily  employs axion monodromy~\cite{Silverstein:2008sg} from massive 4-forms~\cite{Kaloper:2008fb,Kaloper:2011jz} to realize the mechanism of `mutated hybrid inflation'~\cite{Stewart:1994pt}. If we loosen the notion of hybrid inflation to include the smooth sense of `smooth hybrid inflation'~\cite{Lazarides:1995vr} (which otherwise mostly resembles two-field hill-top inflation), then there is also~\cite{Daido:2017wwb} containing a field theory realization of smooth hybrid inflation involving two axions, as well as a regime of aligned inflation reminiscent of hybrid dynamics in~\cite{Peloso:2015dsa}.

			As an immediate benefit, the limited field range $ \Delta\phi_{60} \lesssim\mpl $ implies that harmonic hybrid inflation works with sub-Planckian axion decay constants $ f_a\lesssim\mpl $. This renders our realization of hybrid inflation with axions consistent with bounds on the axion decay constant from controlled string compactifications~\cite{Banks:2003sx,Svrcek:2006yi} as well as from arguments about weak gravity (WGC)~\cite{ArkaniHamed:2006dz}. 
			
			Moreover, the cosine-form of the scalar potential in our model leads to a limited violation of the Lyth bound which generically relates the slow-roll field range and the primordial gravitational wave power measured by the tensor-to-scalar ratio $r$, as already noticed~\cite{Hebecker:2013zda} in models of hybrid natural inflation involving a single axion~\cite{Ross:2009hg}. We will see, that harmonic hybrid inflation is capable of producing a tensor to scalar ratio $r=10^{-4}\ldots 10^{-3}$. However, we find in addition that the inclusion of breaking the $\mathbb{Z}_2$-vacuum degeneracy of harmonic hybrid inflation is necessary to avoid domain wall problems. For the otherwise minimally tuned setup this widens the predicted range of the inflationary observables such that our mechanism generates primordial tensor modes in the range $10^{-4}\lesssim r \lesssim 0.01$. Our string-inspired embedding of hybrid inflation with two axions is thus testable with upcoming CMB B-mode polarization searches (CMB-S3: e.g. Simons Array~\cite{Stebor:2016hgt}, Simons Observatory~\cite{Ade:2018sbj}, Bicep Array~\cite{Hui:2018cvg}; CMB-S4~\cite{Abazajian:2016yjj,Abazajian:2019eic,Abazajian:2019tiv}; and space missions such as e.g. PIXIE~\cite{Kogut:2011xw} or LiteBird~\cite{Hazumi:2019lys,Sugai:2020pjw}).
			
			 Our paper is organized as follows. In section~\ref{sec:singlefield} we consider the effective single field approximation of the model. In section~\ref{sec:phase} we generalize the previous analysis including a non-vanishing phase in one of the cosine terms in the scalar potential, and we evaluate the impact this has on phenomenological predictions. We then discuss the exit from the inflation regime: respectively in section~\ref{sec:preheating} an analysis of the parameters required to meet the experimental constraints is performed, and in section~\ref{sec:Z2break} we address the domain wall problem caused by the $\mathbb{Z}_2$vacuum degeneracy. Based on this we take a look at vacuum stability and regimes of slow-roll eternal inflation in section~\ref{sec:tunnelEI}. In section~\ref{sec:string} we present two possible, qualitatively distinct ways to embed our model in string theory: one way is to consider $C_4$ axions in a Large Volume Scenario (LVS)~\cite{Balasubramanian:2005zx} with three blow-up moduli, the other way is allowing for the presence of magnetized branes in order to generate a potential for $ C_2 $ axions. We conclude by mentioning a possible application of our model to the thraxion scenario described in~\cite{Hebecker:2018yxs}.

	\section{Hybrid inflation with two sub-Planckian axions}
	\label{sec:hybridinflation}
	\subsection{The toy model}
	\label{sec:singlefield}
We start our discussion by looking at the structure of Linde's original hybrid inflation model~\cite{Linde:1993cn} with two scalars $\phi,\chi$. The terms of the scalar potential relevant for the hybrid mechanism read 
\begin{equation}\label{eq:Linde}
V = \frac{\lambda}{4} (\chi^2-v^2)^2+g\chi^2\phi^2+\Delta V(\phi)\quad,
\end{equation}
where $\Delta V(\phi)$ is the non-constant slow roll potential for the inflaton field $\phi$ along $\chi=0$.
Two important features of this model are the presence of two end-of-waterfall minima at $\chi=\pm v$, $\phi=0$ and the bi-quadratic coupling that provides stabilization of $\chi=0$ beyond the waterfall critical point of $\phi$.

With this as guidance, we can guess a two-field axion-like potential of the form
\begin{equation}
V=\Lambda^4+\tilde \Lambda^4 - \left(\tilde \Lambda^4+ \Lambda^4 \cos\left({c}_{1}\,\phi_1\right)\right)\cos\left({c}_{2}\,\phi_2\right),
\label{potential}
\end{equation}
where $\Lambda$ and $\tilde \Lambda$ are energy scales to be specified later (with $\tilde \Lambda>\Lambda$) and the coefficients $c_{i}\geq1$ are proportional to the inverse of the axion decay constants $f_i$ in units of $\mpl$. We take the two fields $\phi_1$ and $\phi_2$ such that their kinetic terms are canonically normalized. Note that $\phi_2$ plays the role of the inflaton field and determines the mass of $\phi_1$. As $\phi_2$ evolves in time, its cosine eventually flips sign and renders $\phi_1$ tachyonic. This is exactly the dynamic of the classic hybrid inflation model. 

At this point it is instructive to perform a `backward' comparison of our axion model with the original hybrid inflation model. In eq.~\eqref{eq:Linde} the potential is given up to quartic terms in the field $\chi$ as well as quadratic in $\phi$, and the structure of the potential covers both the position of the hybrid inflation valley and the minima. We can now expand the axion potential eq.~\eqref{potential} up to quartic order in the fields $\phi_1,\phi_2$, assuming for simplicity $\tilde\Lambda^4=\Lambda^4$. If we do this expansion around $c_1\phi_1=\pi,\phi_2=0$, the resulting scalar potential resembles eq.~\eqref{eq:Linde} qualitatively -- but the two waterfall minima occur at values which are ${\cal O}(1)$ shifted from their position at ($c_1\phi_1=\{0,2\pi\}$ , $\phi_2=0$) in the full potential eq.~\eqref{potential}. We conclude, that the higher-order terms of the `harmonic' cosine terms -- dictated by the instanton expansion -- are crucial for the full field space structure of the model, which we are thus motivated to call `harmonic hybrid inflation'.~\footnote{We note, that in general finding the global minimum and at least a subset of all critical points of such harmonic potentials is non-trivial but achievable using the methods described in Appendix~\ref{app:criticalpoints}.}

We can recast the scalar potential (\ref{potential}) in the canonical form
\begin{equation}\label{finalpotential}
\begin{split}
V=\,&\Lambda^4_1[1- \cos\left({c}_{1}\phi_1+{c}_{2}\phi_2\right)]\\&+ \Lambda^4_2[1-\cos\left({c}_{1}\phi_1-{c}_{2}\phi_2\right)]\\&+ \Lambda^4_3[1-\cos\left({c}_{2}\phi_2\right)].
\end{split}
\end{equation}
by a suitable identification of the parameters, namely
\begin{equation}
\begin{aligned}
\Lambda_1^4&=\Lambda_2^4\equiv \Lambda^4/2\quad, \\
\Lambda_3^4&\equiv\tilde\Lambda^4\quad .
\end{aligned}
\end{equation}
Clearly, the first line is a tuning condition to be fulfilled by any UV realization of the mechanism.

For practical use, we define the following quantities
	\begin{equation}
	V_0 =\tilde\Lambda^4+\Lambda^4 \;,\; \alpha =\frac{\tilde\Lambda^4-\Lambda^4}{\tilde\Lambda^4+\Lambda^4},\  \mbox{and} \ c =c_{2}\quad.
	\end{equation}
	
In a hybrid inflation model, we require the following three conditions on the parameters to be satisfied:
	\begin{enumerate}
	\item Presence of a de Sitter saddle point in the potential. Inflation will start close to this saddle point in order to inflate a sufficient number of e-folds. This amounts to 
	\begin{equation}
	\tilde\Lambda> \Lambda\ .
	\end{equation}
	If this condition is violated, a local minimum will develop that traps the inflaton field instead. 
	\item The dominance of vacuum energy
	\begin{equation}
	\alpha\ll1\ .
	\label{eq:alpha}
	\end{equation}
	This allows for inflation with sub-Planckian field displacements, without the same amount of fine tuning of the initial conditions as usually required in natural inflation.
	\item The inflationary solution should undergo a fast `waterfall transition'. To achieve this, we additionally assume $c_{1} \gg 1$. This drives $\phi_1$ much more massive than $H$ inside the valley for values larger than the `waterfall critical point' $\phi_{2,cr}=\pi/(2c)$, but it becomes strongly tachyonic $-m^2_{\phi_1}/H^2 \gg 1$ after the inflationary trajectory crosses this point.
	\end{enumerate}
	
Between the de Sitter saddle point and the waterfall critical point the waterfall field is stabilized at $\phi_1=\pi/c_{1}$ and, after integrating out $\phi_1$, the scalar potential takes the effective form
	\begin{equation}
	V_\text{inf}(\phi_2)=V_0\left(1-\alpha\cos(c\, \phi_2)\right).
	\label{effscalarpot}
	\end{equation}	

	\begin{figure}[t!]
		\centering
		\includegraphics[scale=0.24]{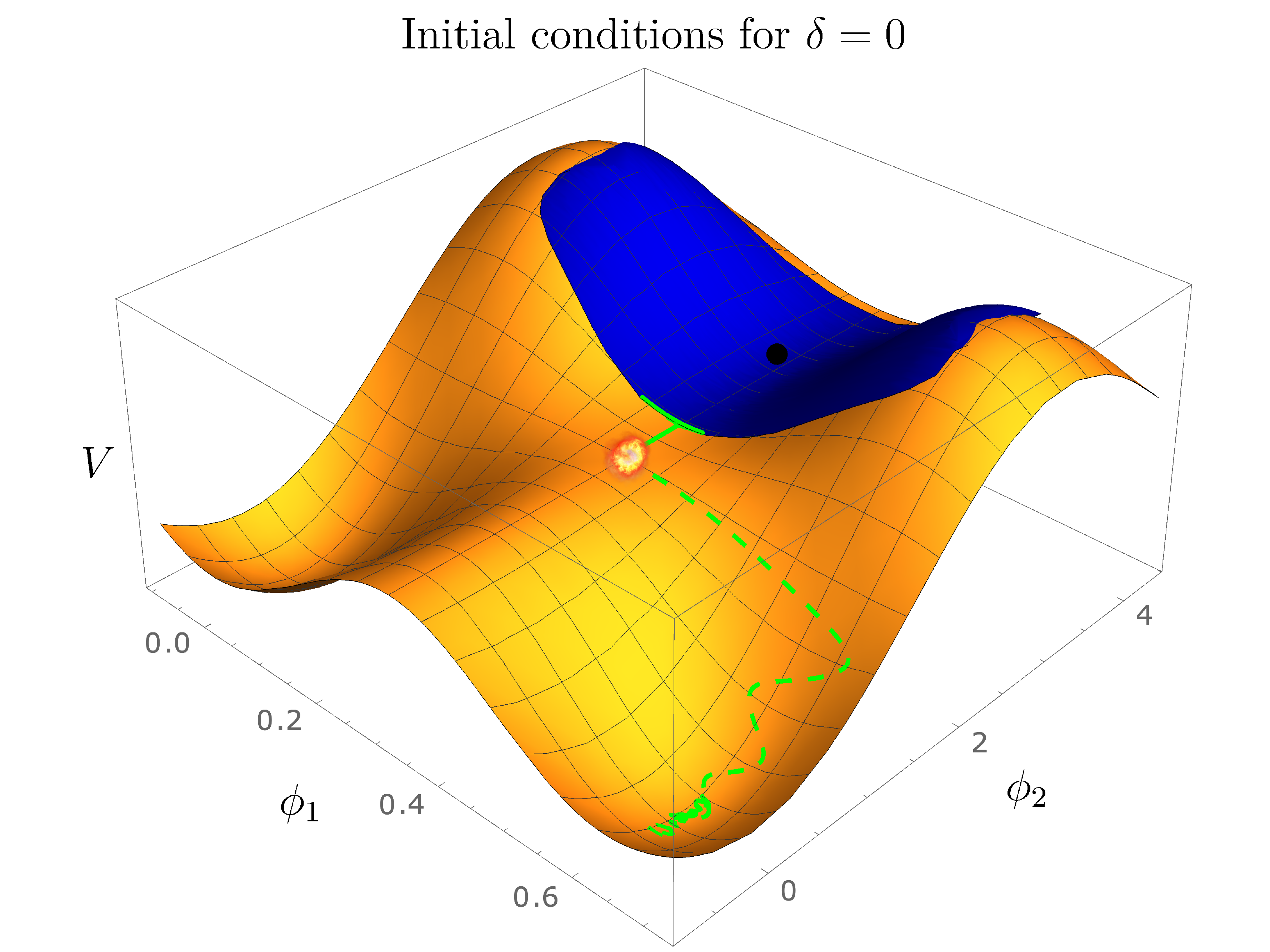}
		\caption{Hybrid inflation potential driven by the two axions $\phi_1$ and $\phi_2$, for the choice $c_{2}=1$, $c_{1} = 10$, $\alpha=0.01$, corresponding to $\tilde\Lambda^4=\frac{1+\alpha}{1-\alpha}\times \Lambda^4\simeq 1.02 \Lambda^4$. The slow-roll regime is along the valley parametrized by the $\phi_2$ direction, while in the $\phi_1$ direction the waterfall is displayed. Black dot: the inflationary saddle point. Blue region: complete region of initial conditions supporting at minimum 60 e-folds of slow-roll inflation. There is no significant initial condition fine-tuning present in this model. Solid green: slow-roll part of a sample inflationary trajectory providing 60 e-folds of slow-roll inflation before reaching the waterfall critical point $\phi_{2,cr}$. Fire ball: explosive growth of tachyonic quantum fluctuations with loss of classical rolling description. Dashed green: would-be classical waterfall evolution neglecting quantum fluctuations for the same sample inflationary trajectory after crossing the waterfall critical point $\phi_{2,cr}$.}
		\label{fig1}
	\end{figure}

We evaluate the viability of our model by computing the slow-roll parameters and the inflationary observables. The slow-roll parameters measure the deviation from an exact de Sitter solution and are defined as
\begin{subequations}
\begin{align}
\epsilon &\equiv -\dfrac{\partial \log H}{\partial N},\\
\eta &\equiv -\dfrac{1}{2}\dfrac{\partial \log \epsilon}{\partial N},
\end{align}
\end{subequations}
i.e. the first slow-roll parameter $\epsilon$ measures the relative change of the Hubble parameter in one expansion time of the universe. Similarly, the second slow-roll parameter $\eta$ captures the relative change of $\epsilon$. The number of e-folds $N$ can be expressed in cosmic time via the relation $dN = Hdt$.

In the slow-roll and single-field regime they can be approximated in terms of derivatives of the scalar potential as follows
\begin{subequations}
		\begin{align}
	 &\begin{aligned}
		\epsilon &\equiv -\frac{\dot H}{H^2}  = \frac{1}{2}\frac{\dot\phi^2}{H^2}\approx \frac{1}{2}\left(\frac{V_\text{inf}^\prime}{V_\text{inf}}\right)^2 = \\ &=\frac{1}{2}\alpha^2 c^2 \sin^2\left(c\, \phi_2\right) + \mathcal{O}(\alpha^4) \label{eps}\ , 
		\end{aligned}\\
		& \eta \equiv -\frac{\dot \epsilon}{2\epsilon H}\approx \frac{V_\text{inf}^{\prime\prime}}{V_\text{inf}} -\left(\frac{V_\text{inf}^\prime}{V_\text{inf}}\right)^2 \label{eta}\ ,
		\end{align}
\end{subequations}
where all the energy scales are in units of $\mpl$ and a prime denotes a derivative with respect to $\phi_2$.

The number $\Delta N$ of inflationary e-folds away from the waterfall critical point $\phi_{2, c}$ can be derived from
	\begin{equation}
	\Delta N\equiv \int_{t}^{t_c} H dt = \int^{\phi_{2,cr}}_{\phi_2} \frac{H}{\dot \phi_2}d\phi_2  = -\int_{\phi_{2,cr}}^{\phi_2} \frac{d\phi_2}{\sqrt{2\epsilon}}\ ,
	\end{equation}
	where in the last step one needs to employ the second Friedman equation.
By performing the integral and inverting the resulting expression we find a relation between the field $\phi_2$ and the number of e-folds
	\begin{equation}\label{phi2Ne}
	\phi_2(\Delta N) = \frac{2\arctan\left(e^{\alpha c^2 \Delta N}\right)}{c}\ ,
	\end{equation}
discarding higher order corrections in $\alpha$. Here we assume we inflate from a point close to the saddle point $\phi_2 = \pi/c$ towards the waterfall critical point $\phi_{2,cr} = \pi/2c$ at $\Delta N =0$, but equivalently inflation could start close to any of the other saddle points $\phi_2 = (\pi+2\pi n)/c$ towards the transition points $\phi_{2,cr} = (2\pi n \pm \pi/2)/c$. The above solution allows us to write the first slow-roll parameter as a function of e-folds
\begin{equation}
\epsilon(\Delta N) = 2 \alpha \gamma \frac{ e^{2 \gamma \Delta N}}{ \left(1+e^{2 \gamma \Delta N}\right)^2},
\label{eq:eps0}
\end{equation}
where we define 
\begin{equation}
\gamma \equiv \alpha c^2\quad.
\label{eq:gammadefinition}
\end{equation} 

\begin{figure}[t!]
		\centering
		\includegraphics[scale=0.22]{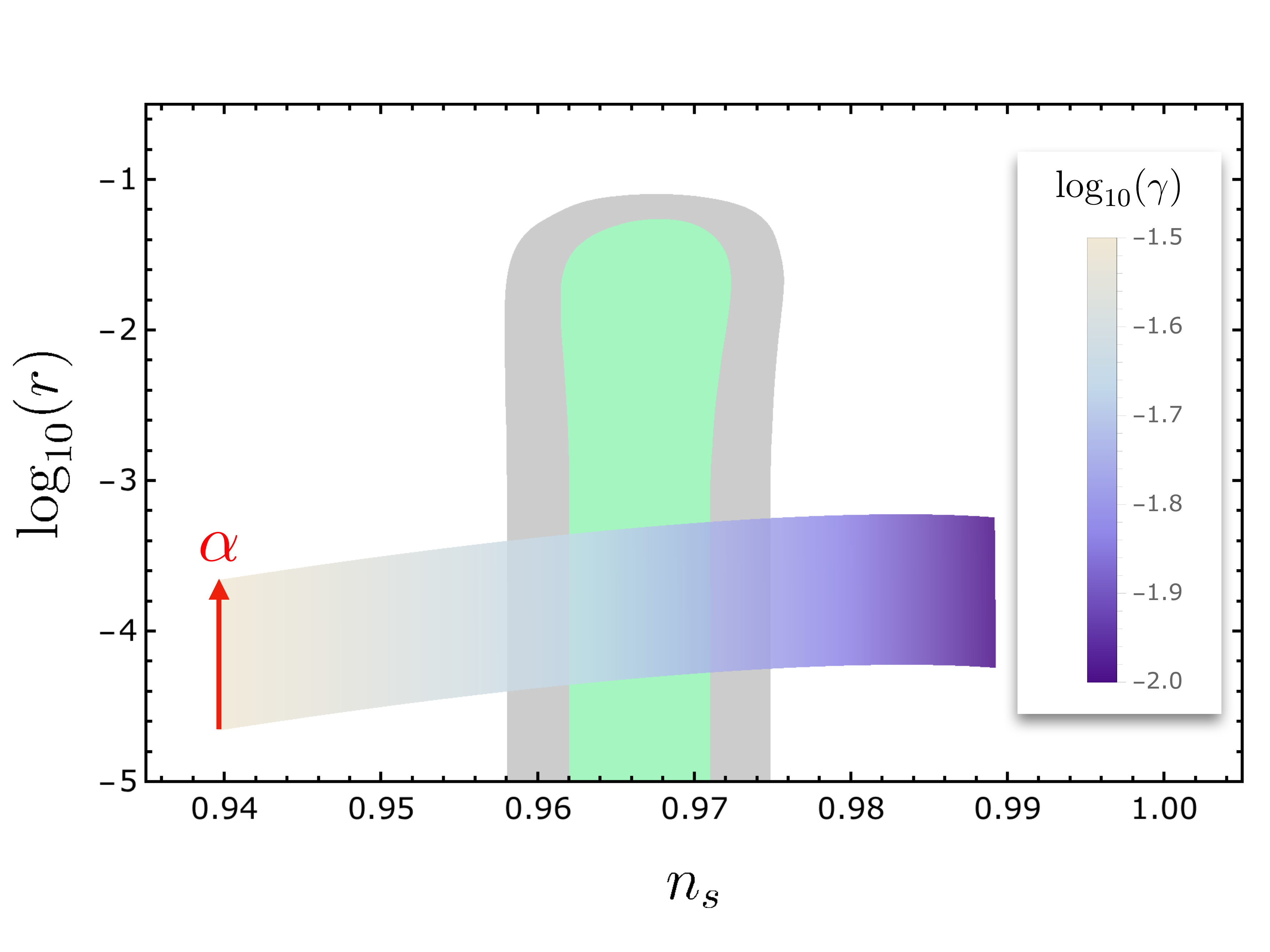}
		\caption{We show the analytical predictions for $n_s$ and $r$ varying 	$\log_{10}(\gamma) \in [-3,-1.5]$ and $\log_{10}(\alpha) \in [-3,-2]$ (purple-to-white contour) together with the $1\sigma$ and $2\sigma$ confidence contours found by \textit{Planck}~\cite{Akrami:2018odb} (grey and mint contours, respectively), where they combined the \textit{Planck} data the with \textit{BICEP2/Keck Array}~\cite{Ade:2018gkx} and BAO~\cite{Beutler:2011hx, Ross:2014qpa,Alam:2016hwk} data.}
		\label{fig:nsrgammaalpha}
\end{figure}

Moreover, the second slow-roll parameter is then given by
\begin{equation}
\eta(\Delta N) \equiv \frac{1}{2 \epsilon} \frac{\partial \epsilon}{ \partial \Delta N} =  \gamma \frac{1 - e^{2 \gamma \Delta N}}{1+ e^{2 \gamma \Delta N}}\ .
\label{eq:eta0}
\end{equation}
Notice the hierarchy $|\eta| \gg \epsilon$. 
The tensor-to-scalar ratio and the spectral tilt are given by 
\begin{equation}
\begin{aligned}
r&= 16\epsilon \, ,\\
n_s&=1-2\epsilon+2\eta\, .
\end{aligned}
\end{equation}
Therefore, as usual in small field inflation, the tensor-to-scalar ratio is highly suppressed. The spectral tilt is to a high level of precision determined by the value of $\gamma$ only (given some value of $\Delta N$). If we take $\Delta N$ between 50 and 60 e-folds the $1\sigma$ constraint that $n_s \in [0.9627, 0.9703]$~\cite{Akrami:2018odb} translates into $\gamma \in [0.0185 , 0.0229]$. In other words, the parameters $\Lambda$, $\tilde \Lambda$ and $c$ are degenerate, however, they need to be fine-tuned such that the resulting $n_s$ falls within the observed window.
As an explicit example~\ref{fig1}, if we choose the following values for the parameters, $\alpha=0.02$ and $c=1$, this model predicts
\begin{equation*}
\mbox{for } \Delta N=50 : \,\, \left\lbrace
\begin{aligned}
r&= 0.0013\\
n_s&=0.969504
\end{aligned}\right.
\end{equation*}

\begin{equation*}
\mbox{for } \Delta N=60 : \,\, \left\lbrace
\begin{aligned}
r&= 0.00094\\
n_s&=0.966847
\end{aligned}\right.
\end{equation*}
More generally, in Fig.~\ref{fig:nsrgammaalpha} we show the analytical predictions for $n_s$ and $r$ based on eq.s~\eqref{eq:eps0} and \eqref{eq:eta0} with $\Delta N = 60$, varying $\log_{10}(\gamma) \in [-3,-1.5]$ and $\log_{10}(\alpha) \in [-3,-2]$. As discussed before $\alpha$ only moves the prediction for $r$ up or down and $n_s$ is very sensitive to the value of $\gamma$. 
As we will consider in a moment, we have to break the symmetry in $\phi_1$ in order to avoid the formation of domain walls. Therefore, this toy model is not the end of the story and we will see that the predictions are also highly sensitive to the amount of symmetry breaking.

\subsection{Effect of instanton-induced phases}
\label{sec:phase}
Since axions arising in string theory enjoy perturbative shift symmetries as non-linear realizations of the underlying 10D p-form gauge symmetries, axions acquire a scalar potential via non-perturbative instanton effects (unless monodromy-generating sources of stress-energy such as branes or fluxes are present as well). These instanton effects generating the scales $\Lambda_i^4$ of the periodic axion potentials they induce, possess an ab initio arbitrary complex phase. In string theory realizations the 1-loop determinants of such instanton contributions, entering the scales $\Lambda_i^4$, become functions of moduli VEVs. Since we can tune the VEVs by the choice of quantized background fluxes of p-form field strengths, the value of the phases of the instanton effects is adjustable. Hence, in principle all of the three cosines in eq.~\eqref{finalpotential} can have a non-vanishing but finite adjustable phase, which was omitted in the discussion so far. However, two phases out of three can be reabsorbed thanks to the shift freedom given by the presence of two axions. In the following, we evaluate how much the inclusion of the remaining phase will change the model and how it will affect the inflationary predictions. 

Without loss of generality, we choose to keep the phase $\vartheta$ in the single-axion cosine term. Again, during the slow-roll evolution of $\phi_2$ we can integrate out $\phi_1$. This leaves us with   
\begin{equation}\label{1axionpotential}
	V_\text{inf}^{(\vartheta)}(\phi_2)=V_0-\tilde{\Lambda}^{4} \cos\left(c_{2}\phi_2+\vartheta\right)+\Lambda^{4}\cos\left(c_{2}\phi_2\right),
	\end{equation}
and the equations determining the slow-roll parameters change accordingly. In fact, the first slow-roll parameter becomes to leading order in $\alpha$ and $\vartheta$ 


\begin{equation}\label{etheta}
\begin{aligned}
	\epsilon^{(\vartheta)} \simeq & \frac{c_{2}^2}{2}\left( \alpha \sin(c_{2}\phi_2) + \frac{1}{2}\vartheta \cos(c_{2} \phi_2)\right)^2 + \\
	 & \qquad\qquad\qquad + \mathcal{O}(\alpha^2 \vartheta,\alpha^2 \vartheta^2)\quad .
\end{aligned}
	\end{equation}

	\begin{figure*}
		\centering
		\includegraphics[scale=0.20]{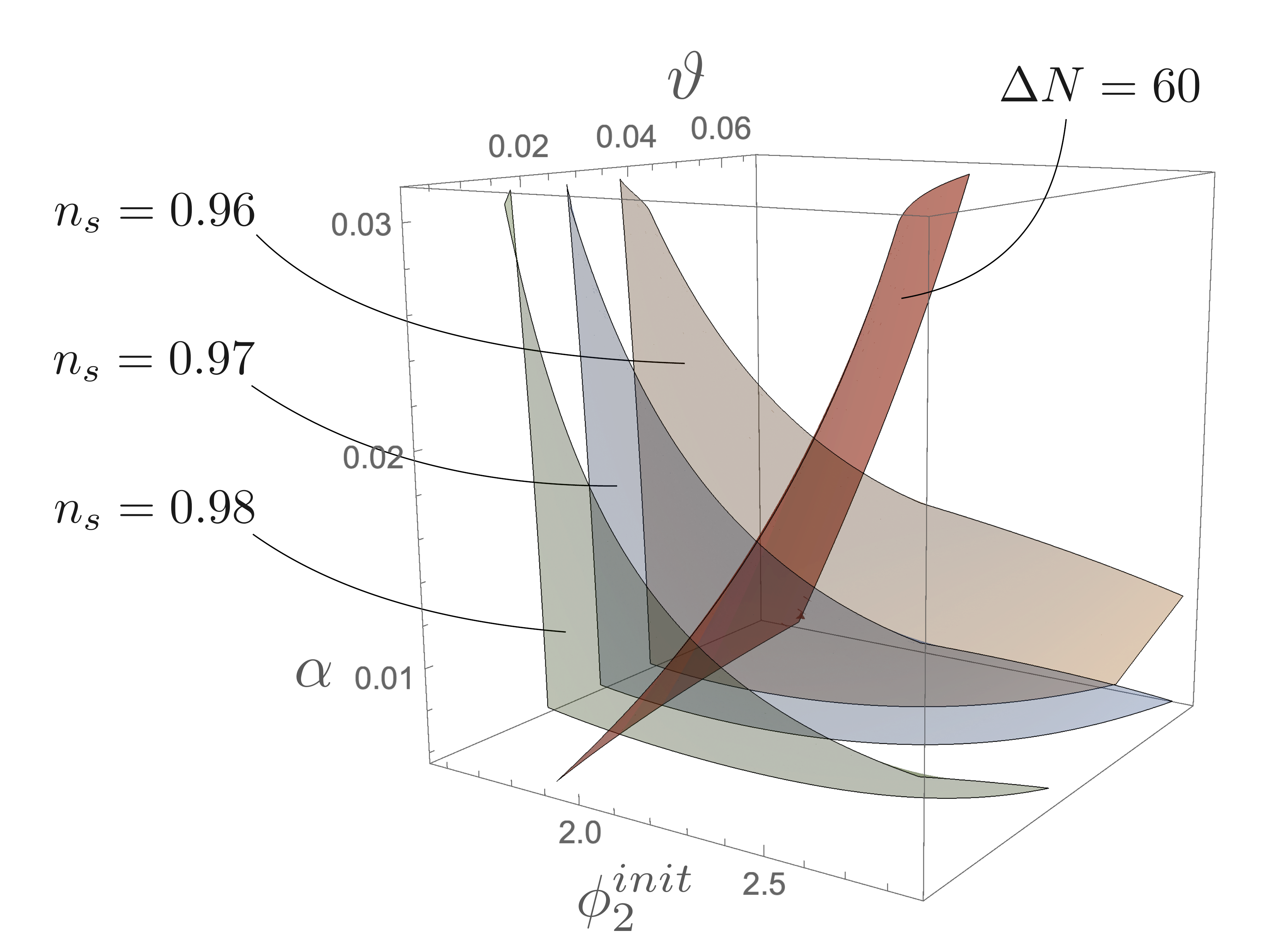}
		\includegraphics[scale=0.18]{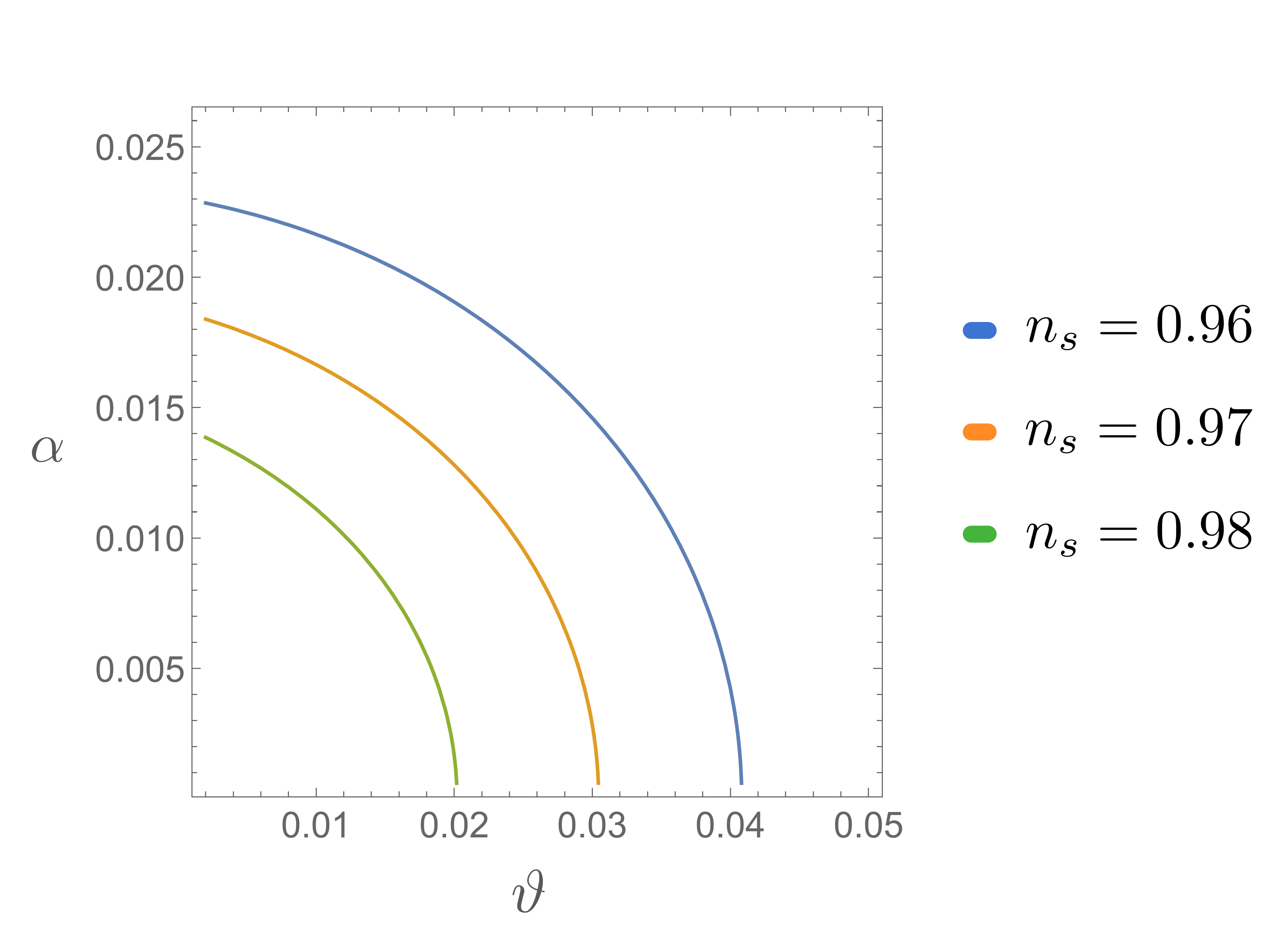}
		\includegraphics[scale=0.23]{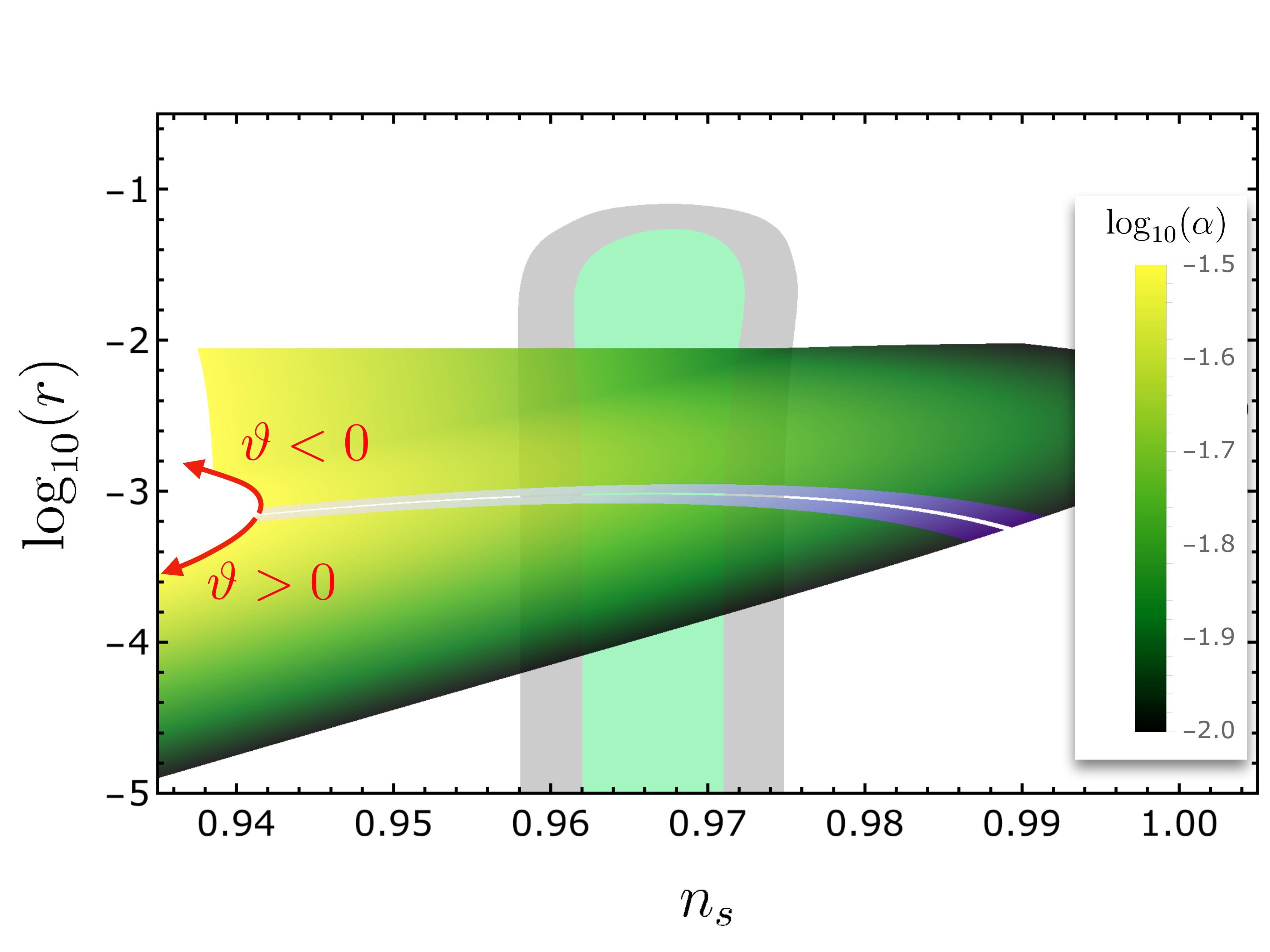}
		\caption{Regions in parameter space spanned by the two harmonic hybrid inflation potential and resulting predictions for $n_s$ and $r$. We evaluated which values of the parameters $\alpha$, $\phi_2$ (meaning its starting point on the cosine) and especially $\vartheta$ fit best the experimental data. In the first plot, the three surfaces are the regions of parameter space which give the three values of $n_s$ that are displayed; the red one is the space spanned by the values of the parameters that give 60 e-folds. The intersections of this with the other three surfaces give the best choices of parameters that can reproduce the experimental data. In the second plot, we show more precisely the interplay between $\alpha$ and $\vartheta$ at a given spectral tilt. In the third plot, we compare the \textit{Planck} contours with the numerical predictions for $ n_s $ and $ r $ keeping $ c=1 $ fixed while varying $\log_{10}(\alpha)\in \left[-2.0,-1.5\right]$ and $\vartheta\in \left[-1/15,1/15\right]$. Notice that here we allow the phase to take both positive and negative values. The purple contour is the region where the analytical approximation for small $\vartheta$ holds. This will be useful to compare the modifications from the perfect hybrid due to $\vartheta$ with the ones following from the $\mathbb{Z}_2$-symmetry breaking terms (see ~\ref{sec:Z2break})} 
		\label{fig3}
	\end{figure*}

We require $\vartheta\lesssim 0.1$ in order to have vacuum energy domination during the slow-roll regime. Otherwise, for bigger values of $\vartheta$ the hybrid mechanism is spoiled, because the inflaton-dependent part of the scalar potential controls the inflaton dynamics, which in turn drives the model into the large-field regime. The results of the analysis for the inclusion of the phase are displayed in Fig.~\ref{fig3}. There is a whole set of ($\alpha$, $\vartheta$, $\phi_2$) combinations that can actually give good hybrid inflation lasting (at least) 60 e-folds.  Notice that below a certain value of $\alpha$, the value of the phase giving the required $ n_s$ becomes basically fixed. Thus, one could balance the values of $\alpha$ and $\vartheta$ in order to have the least amount of fine-tuning possible.~\footnote{We wish to point out in passing, that it may be possible to get the instanton phases discussed in this section to vanish dynamically in appropriately arranged field theory realizations involving bi- or tri-fundamental matter representations coupled to three non-Abelian gauge groups. We thank N. Kaloper for illuminating discussions on this point, but leave this possibility and the potentially stringent constraints imposed by string theory embeddings of such setups for future work.}

\subsection{Preheating and domain walls}
	\label{sec:preheating}
At the end of inflation the hybrid valley false vacuum has to decay into the true vacuum. The onset of this transition is controlled by the inflaton field, which allows for tachyonic growth of the waterfall field as the inflaton crosses the critical point.  Tachyonic preheating proceeds similar to a second order phase transition, via a spinodal decomposition. The spinodal time and the average size of the domains have been computed originally in~\cite{Felder:2000hj,Felder:2001kt} and have been refined beyond the quench approximation in~\cite{Asaka:2001ez,Copeland:2002ku,Lyth:2010zq}. However, in order to avoid the domain wall problem~\cite{Zeldovich:1974uw} either the tachyonic instability should be avoided or the vacuum degeneracy needs to be broken such that asymmetry generates a pressure pushing the true vacuum domains to grow with a rate that depends on the surface tension of the domain walls.
For the latter option we need $\Delta V \gg V_0 \pi^4/ c_1^2$ such that domain walls will not dominate the energy budget in the universe~\cite{Vilenkin:1984ib}. 
In order to break the vacuum degeneracy we necessarily need to break the $\mathbb{Z}_2$ symmetry around the inflationary trajectory. Our model naturally allows for such a breaking by assuming different axion decay constants multiplying $\phi_1$. We investigate this simple generalization in section~\ref{sec:Z2break}. Additionally, such a setup \sout{generically} removes the tachyonic instability and hence no domain walls are formed in the first place. This drastically changes the physics of preheating, as it does not proceed via the spinodal instability. Instead, a period of parametric resonance~\cite{Mukhanov:2005sc, Kofman:1994rk, Kofman:1997yn} will most likely follow and the post-inflation phenomenology may be enriched by the formation of two-field oscillons~\cite{Gleiser:2011xj}, an exciting possibility we wish to investigate in more detail in future work.

\subsection{Breaking the $\mathbb{Z}_2$ vacuum degeneracy}
	\label{sec:Z2break}
As discussed in the previous section, in our toy model of harmonic hybrid inflation, the two vacua  providing the possible endpoints of the waterfall regime have a $\mathbb{Z}_2$-symmetry (in the $\phi_1$ direction). This leads to the formation of domain walls after inflation. Therefore, we should break the symmetry to avoid such a scenario incompatible with our universe. 

\subsubsection{Summary of $\mathbb{Z}_2$ symmetry breaking effects}
	\label{sec:Z2breakSummary}

\begin{figure}[t!]
		\centering
		\includegraphics[scale=0.25]{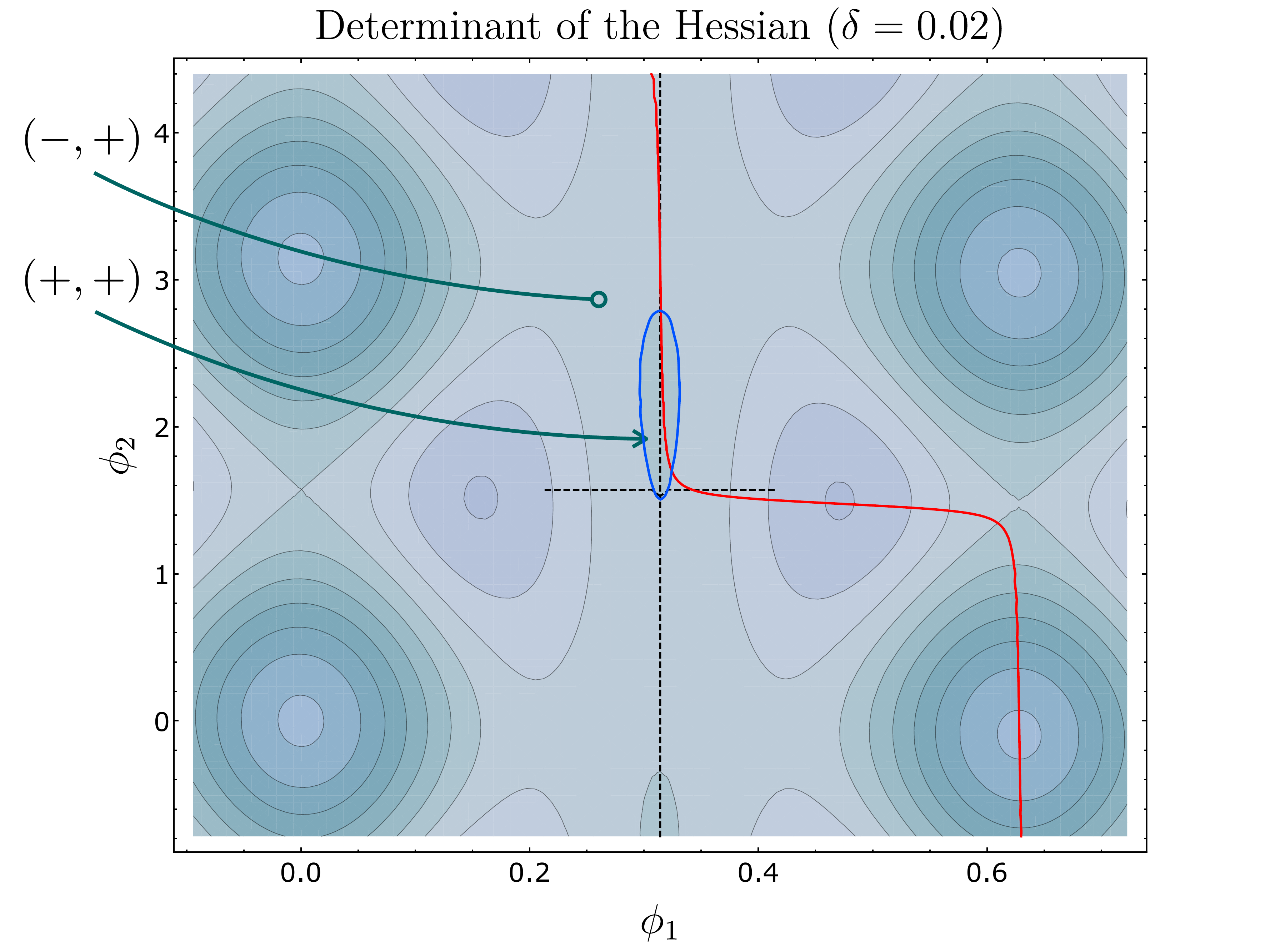}
		\caption{Contour plot of the determinant of the Hessian of the deformed potential given in eq. \ref{fullpotential} with $c_{1}^\pm=c_{1}(1\pm\delta)$ and $c_{2}^\pm = c$.  We choose the same values of the parameters as in Fig.~\ref{fig:ICSwithtrajectories}, namely $\delta = 0.02$, $\alpha = 0.01$, $\gamma = 0.011$ and $c_1 = 10$. The red solid curve shows where the gradient of the potential in the $\phi_1$ direction is vanishing, and provides initially a good proxy of the inflationary trajectory. The black cross, on the other hand, shows the vanishing of the gradient in case the $\mathbb{Z}_2$ symmetry would be unbroken, that is, for $\delta =0$ instead. The highlighted central contour specifies where the determinant of the Hessian changes sign. This contour shrinks to zero size as $\delta \rightarrow 0$, while the rest of the contours remain similar. }
		\label{fig:hessian}
	\end{figure}

The simplest way to do so is to generalize the potential eq.~\eqref{finalpotential} to the form
\begin{equation}\label{fullpotential}
\begin{split}
 V=\,&\Lambda^4_1[1- \cos\left({c}_{1}^+ \phi_1+{c}_{2}^+\phi_2\right)]\\&+ \Lambda^4_2[1-\cos\left({c}_{1}^-\phi_1-{c}_{2}^-\phi_2\right)]\\&+ \Lambda^4_3[1-\cos\left({c}_{2}\phi_2\right)].
 \end{split}
 \end{equation}
 Choosing $c_{1}^+\neq c_{1}^-$ will break the $\mathbb{Z}_2$ vacuum degeneracy, but it also removes the tachyonic instability at the end of inflation and no domain walls are formed. Moreover, preheating does not proceed via tachyonic growth. (Assuming that $c_1^+-c_1^-$ is not exponentially close to zero, which we expect to be the case in the possible string embeddings that are discussed in section~\ref{sec:string}). To see why the tachyonic instability characteristic for tachyonic preheating disappears, we notice that the inflationary trajectory proceeds along a straight line $\phi_1 = \pi/c_1$ only in the presence of the $\mathbb{Z}_2$  symmetry. When the $\mathbb{Z}_2$ symmetry is broken, the waterfall field $\phi_1$ gets stabilized at a value depending on the expectation value of the inflaton field $\phi_2$. This is illustrated with the red solid line in Fig.~\ref{fig:hessian}.  More precisely, in presence of the unbroken  $\mathbb{Z}_2$ symmetry, the signature of the Hessian changes along the inflationary slow-roll trajectory $\phi_1=\pi/c_1$ from $(-,+)$ to $(+,-)$ at a single point $(\phi_1=\pi/c_1,\phi_2=\phi_{2,cr})$. Once the $\mathbb{Z}_2$ symmetry is broken, we see in Fig.~\ref{fig:hessian} that the path given by $\partial_{\phi_1}V=0$ approximating the slow-roll part of the inflaton trajectory enters a region with Hessian of signature $(+,+)$ while turning away and missing the former critical point before reentering a region with signature $(-,+)$. From this it is self-evident that the point like transition from $(+,-)$ to $(-,+)$ signature on a classical trajectory with $\phi_1=const.$ characteristic of the tachyonic preheating instability is simply gone once the $\mathbb{Z}_2$ symmetry is broken.  This means $\phi_1$ never develops the tachyonic preheating instability, except in the presence of a $\mathbb{Z}_2$  symmetry.
 
 We will look at a minimal deformation of this kind, choosing a symmetric deformation $c_{1}^\pm=c_{1}(1\pm\delta)$ while keeping $c_{2}^\pm=c$, $\Lambda_1^4=\Lambda_2^4 = \Lambda^4/2$ and $\Lambda^4_3 = \tilde \Lambda^4$, which should be close but slightly larger than $\Lambda^4$ (see the discussion around eq.~ \ref{eq:alpha}).  Notice that the effect of symmetry breaking becomes larger if we move further away from $\phi_1 =0$. At the same time, a set of global minima are still located at $\phi_1 =0$ and $\phi_2 = 2\pi n /c\,$.  From here onwards we restrict ourselves to inflation in the neighbourhoods of these minima.

 We shall now quickly summarize the effects of this $\mathbb{Z}_2$-symmetry breaking before providing the details. Breaking the degeneracy in the minimal fashion described above has consequences. First, the maximum of the potential in $\phi_2$ in the hybrid valley around $\phi_1=\pi/c_{1}$ is shifted from its original position at $\phi_2^{max}(\delta=0)=\pi/c$. Next, the symmetry breaking implies that the `waterfall' regime, involving tachyonic preheating proper as well,  in the strict sense is lost and replaced with a waterfall-like rapid exit from inflation once $\phi_2<\phi_{2,cr}$. Furthermore, the loss of degeneracy implies that there are now two inequivalent inflationary regimes: one starting from $\phi_2<\phi_2^{max}$ towards $\phi_2=0$ and another starting at $\phi_2>\phi_2^{max}$ towards $\phi_2=2\pi/c$ in the neighbouring vacuum region. In addition, with increasing $\delta$ the field trajectory will become more and more curved in its entirety due to the increasingly broken $\mathbb{Z}_2$-symmetry of the inflationary valley, which will cause sizeable changes in $n_s$ and $r$. As we will see in Fig.~\ref{fig:nsrnumerical} and Fig.~\ref{fig:viableparameterspace} using the full numerical analysis, maintaining compatibility with observations forces us to keep $\delta\ll 1$ maintaining a near-hybrid inflation regime. 
 
 \subsubsection{Analytical treatment of $\mathbb{Z}_2$ symmetry breaking effects}
	\label{sec:Z2breakAnalytics}
 
 \begin{figure}[t!]
 	\centering
 	\includegraphics[scale=0.24]{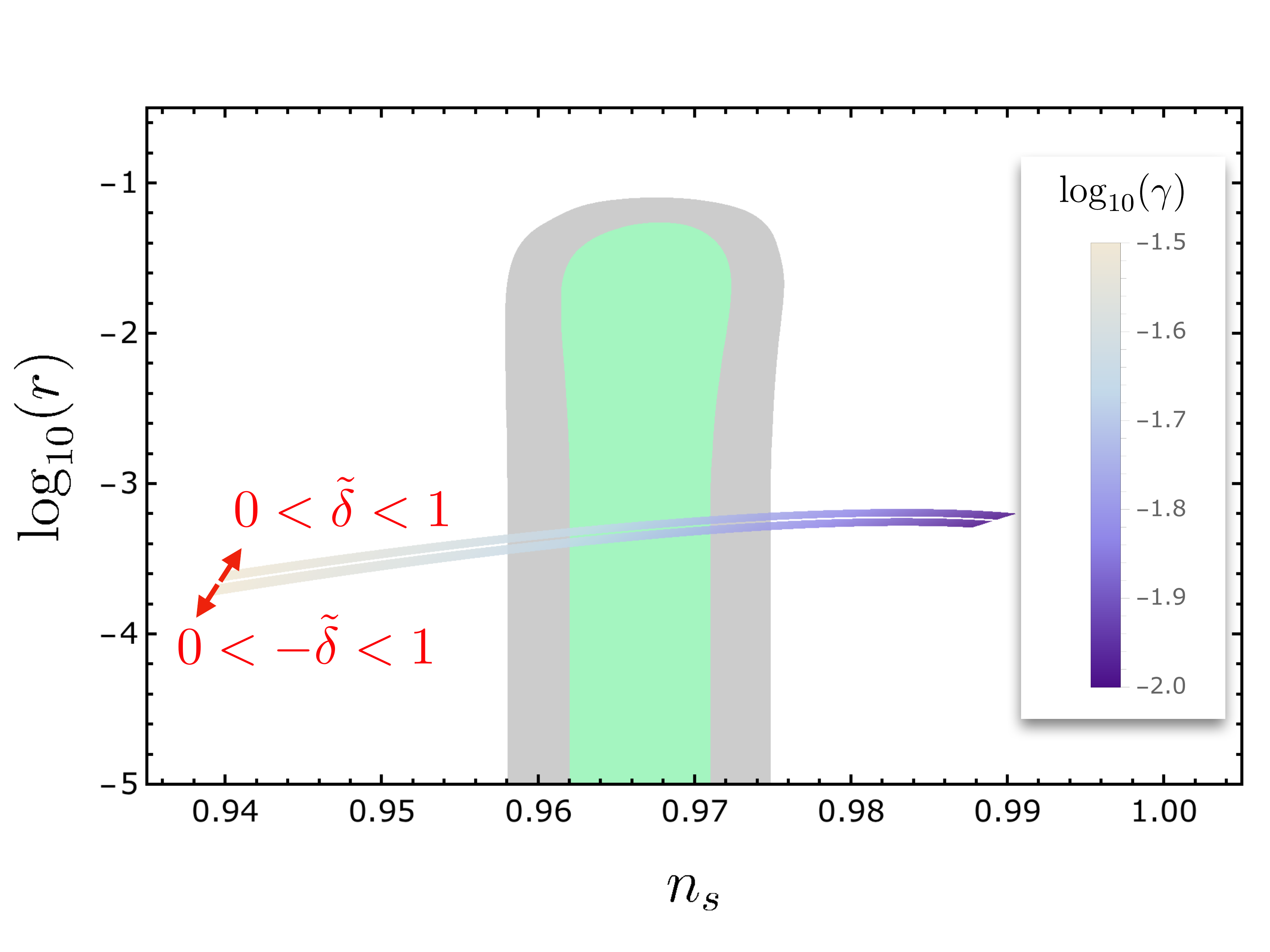}
 	\caption{Same as Fig.~\ref{fig:nsrgammaalpha} but now we compare the \textit{Planck} contours with the analytical predictions for $n_s^{(\delta)}$ and $r^{(\delta)}$ and $\log_{10}(\gamma) \in [-2.0,-1.5]$, where we vary $\log_{10}(|\tilde \delta|) \in [-2.0,-1.0]$ while fixing $\alpha = 0.01$. The region $\tilde\delta >0$ corresponds to inflationary trajectories with wide initial condition range of the type shown in Fig.~\ref{fig:ICSwithtrajectories} ending up in the false minimum at $\phi_1=2\pi/c_1$, $\phi_2=-{\cal O}(\tilde\delta)$.The complement  $\tilde\delta <0$ corresponds to trajectories which in Fig.~\ref{fig:ICSwithtrajectories} start with $\phi_2$-values beyond the saddle point with little allowed initial condition space, and end up in the true minimum at $\phi_1=\phi_2=0$.}
 	\label{fig:nsrgammadelta}
 \end{figure} 

First we study analytically a small breaking of the $\mathbb{Z}_2$ symmetry. We start by expanding the potential to linear order in $c_{1} \phi_1 \delta$. The result is 

\begin{equation}
\begin{split}
 V^{(\delta)}=\,& V^{(0)} +  c_{1} \phi_1 \delta\ \frac{\Lambda^4}{2} \left(\sin_+ -\sin_- \right) + \mathcal{O}\left((c_{1} \phi_1 \delta)^2\right).
 \end{split}
 \end{equation}
Here, $V^{(0)}$ denotes the hybrid potential eq.~\eqref{finalpotential} and we introduced the short-hand notation $\sin_\pm \equiv \sin\left({c}_{1} \phi_1\pm {c}\phi_2\right)$. The higher-order terms in the potential are the ones arising from the series expansion at least quadratic in $c_{1} \phi_1 \delta $, the even terms coming with factors of $\cos_\pm$. Hence, we see that choosing $\delta\neq 0$ breaks the $\mathbb{Z}_2$ vacuum degeneracy between the two vacua left and right of the inflationary valley 
 with an amount of $\Delta V \sim (\pi \delta)^2 \Lambda^4$, allowing us to avoid the domain wall problem. Repeating the same steps as in section \ref{sec:hybridinflation}, the effective potential for $\phi_2$ in the $c_{1} \phi_1= \pm \pi + \mathcal{O}(\delta)$ valley is, to leading order in $\delta$, given by

 \begin{equation}
 V_\text{inf}^{(\delta)} = V_0\left(1 - \alpha \cos(c \phi_2) \mp \alpha \tilde \delta \sin(c \phi_2)  \right) + \mathcal{O}(\pi^2 \delta^2)  ,
 \end{equation}
 
 where we defined $\tilde\delta \equiv \frac{ \Lambda^4  }{\Lambda^4 + \tilde\Lambda^4}\frac{\pi \delta}{\alpha}=\frac{1-\alpha }{2}\frac{\pi \delta}{\alpha}$. Notice that the phenomenology will be close to our hybrid inflationary toy model only if $ |\pi\delta| \ll \alpha$. For larger $\alpha \lesssim |\pi\delta| \ll 1$ the inflationary trajectory might still be of the hybrid kind, in the sense that inflation happens along a straight line $c_{1} \phi_1\approx \pm \pi$ and ends almost instantaneously at a critical value of $\phi_2$. However, the potential is twisted and the saddle points are displaced along the $\phi_2$ direction. This squeezes or stretches the effective potential and is the reason why the phenomenology is substantially modified. We therefore trust the following analytical predictions only in the regime $|\tilde \delta| \ll 1$ and we will study larger deformations numerically. From the effective potential we compute the slow-roll parameter 
 \begin{equation}\label{edelta}
 \epsilon^{(\delta)} \approx \frac{1}{2}c^2 \alpha^2 \left(\sin(c \phi_2) \mp \tilde \delta \cos(c \phi_2) \right)^2\ ,
 \end{equation}
 where we neglected corrections of order $\mathcal{O}(\alpha, \tilde\delta^2)$ \textit{inside} the round brackets. 
 Moreover, using the expression for $V^{(\delta)}$ we find that $\phi_1$ becomes tachyonic at $c \phi^{(\delta)}_{2,cr} = 2 \pi n \pm \arctan\left(\frac{1}{\pi\delta}\right) \approx c \phi_{2,cr}$, i.e. the waterfall transition point is approximately unchanged. Next, by integrating over the slow-roll parameter we express $\phi_2$ as a function of e-folds. Two branches of solutions emerge
 \begin{equation}
		\phi_2^{(\delta)}(\Delta N) = \frac{2 \arctan\left(e^{\gamma \Delta N}\right) }{c} \pm \frac{\tilde \delta}{c} \frac{\left(1 -e^{\gamma \Delta N}\right)^2 }{1 +e^{2\gamma \Delta N}}
\end{equation}
 where as before we defined $\gamma \equiv \alpha c^2$ and we discarded corrections that are of higher order in $\tilde \delta$. Moreover, we assume that we start inflation in the neighbourhood of the saddle point close to $c \phi_2 = \pi$ and move towards the critical point close to $c \phi_2 = \pi/2$. The alternative trajectories starting close to $c \phi_2 = -\pi$ moving towards $c \phi_2 = -\pi/2$ have equivalent phenomenology with these two solutions, but with the $\pm$ sign swapped. In light of the numerical analysis, we therefore cover all possible outcomes by solving for the inflationary solution that starts close to $c \phi_2 =c_{1} \phi_1 = \pi$ where $\delta$ takes both positive and negative values. Therefore, we take $\pm$ to be $+$ from here onwards. The first slow-roll parameter is then given by 
 \begin{equation}
 \epsilon^{(\delta)}(\Delta N) =2\alpha \gamma \frac{ e^{2 \gamma \Delta N}\left(1-\tilde\delta +e^{2 \gamma \Delta N}(1+\tilde \delta)\right)^2}{\left(1+ e^{2 \gamma \Delta N}\right)^4}\ .
 \label{eq:epsdelta}
 \end{equation}
 Moreover, the second slow-roll parameter is, to leading order in $\tilde \delta$, given by
 \begin{equation}
 \eta^{(\delta)}(\Delta N) = \gamma \frac{1-e^{2\gamma \Delta N} }{1+e^{2\gamma \Delta N}} +4 \gamma \tilde \delta \frac{ e^{2\gamma \Delta N}}{\left(1+e^{2\gamma \Delta N}\right)^2}\ ,
 \label{eq:etadelta}
 \end{equation}
 and, as before, provides the dominant contribution to $n_s$. 
 The tensor-to-scalar ratio and the spectral tilt read 
\begin{equation}
	\begin{aligned}
	r^{(\delta)}&= 16\epsilon^{(\delta)},\\
	n_s^{(\delta)}&=1-2\epsilon^{(\delta)}+2\eta^{(\delta)}.
	\end{aligned} 
	\end{equation} 
In Fig.~\ref{fig:nsrgammadelta} we plot the analytical predictions where we fix $\Delta N = 60$ and $\alpha = 0.01$. We vary $\log_{10}(\gamma) \in [-2.0,-1.5]$ and $\log_{10}(|\tilde \delta|) \in [-2.0,-1.0]$. Within the regime of validity of the analytical approximation, the predictions are very sensitive to the value of $\gamma$, and only mildly dependent on $\tilde \delta$.

 \subsubsection{Numerical analysis of $\mathbb{Z}_2$ symmetry breaking effects}
	\label{sec:Z2breakNumerics}	

\begin{figure}[t!]
	\centering
	\includegraphics[scale=0.25]{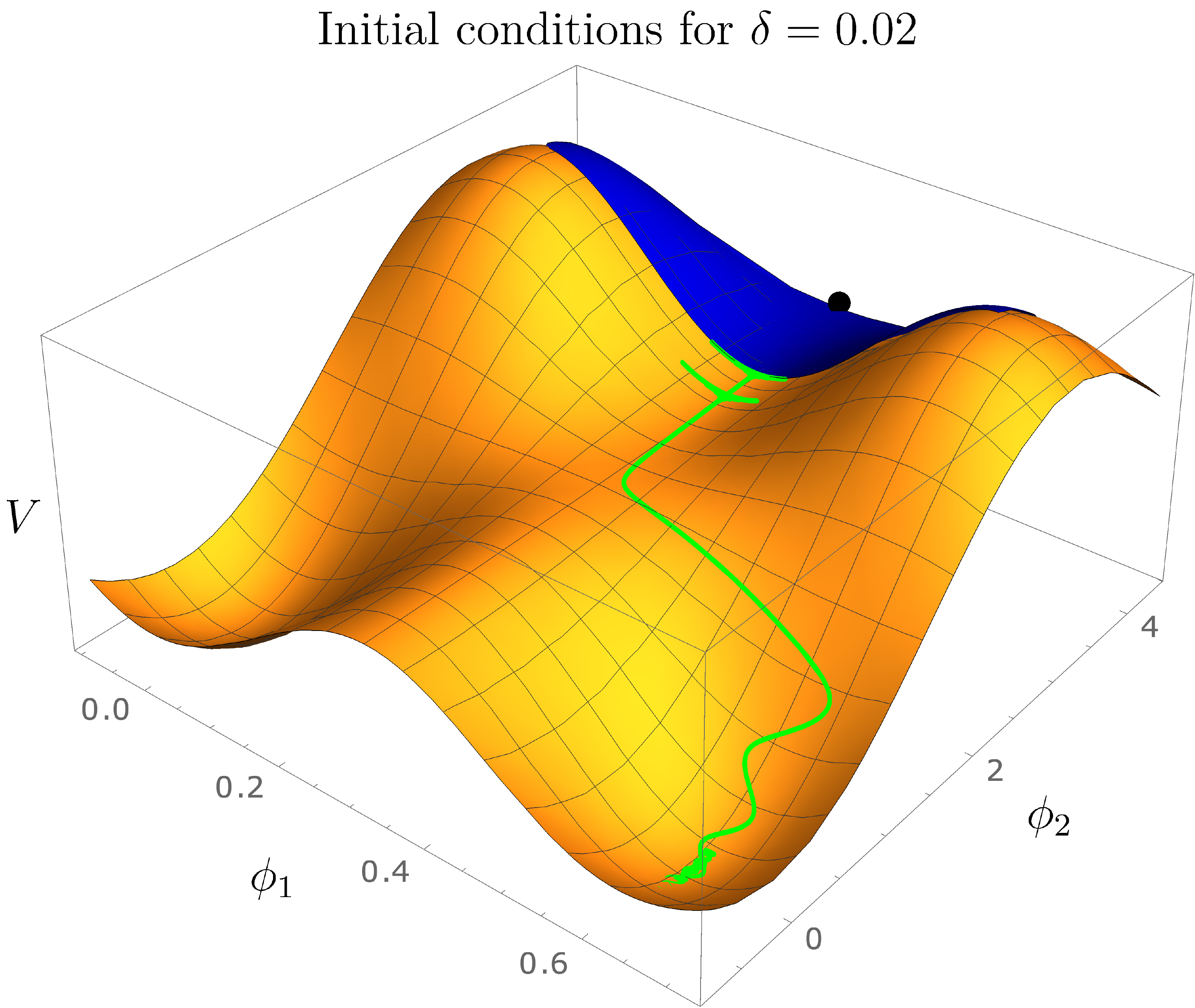}
	\caption{We illustrate the amount of fine-tuning of initial conditions in a deformed hybrid potential with $\delta = 0.02$, $\alpha =0.01$, $\gamma = 0.011$ and $c_{1} = 10$. In blue the patch of the potential around the saddle point (black dot) from which at least 60 e-folds of inflation will originate is shown. Moreover, in green two example trajectories of respectively 50 and 60 e-folds are shown.}
	\label{fig:ICSwithtrajectories}
\end{figure}

Next, we study larger deformations numerically. In order to set appropriate initial conditions we have to identify the location of the saddle point~\footnote{For a discussion of finding critical points in a more general scalar potential that is a sum of cosines (and/or sines), but where the coefficients appearing in front of the fields are integers, see appendix~\ref{app:criticalpoints}.}, which we numerically search for in the neighbourhood of $c_{1}  \phi_1 = \pi$ and $c \phi_2 = \pi$, given some $\delta$. Even though the potential might get twisted substantially, it turns out we do not need much fine-tuning of the initial conditions. In Fig.~\ref{fig:ICSwithtrajectories} we show an example of a deformed hybrid potential with $\delta = 0.02$ and $\alpha =0.01$ (this corresponds to $\tilde\delta \approx 3.1$.) Moreover, we take $\gamma = 0.011$, because, as we will see in a moment, this gives predictions compatible with the CMB constraints. Using the relation $\gamma \equiv \alpha c^2$, this means that $c \approx 1.0$. The blue patch around the saddle point is the region from which at least 60 e-folds of inflation originate. Moreover, two example trajectories of respectively 50 and 60 e-folds are shown. They are separated by $5-10 \%$ of the total field range of $\phi_2$ between the saddle point and the minimum of the potential.   

We solve for the background solution until the end of inflation using the transport code~\cite{Dias:2015rca} and evaluate the Hubble slow-roll parameters at $60$ e-folds before the end of inflation to estimate the tensor-to-scalar ratio and the spectral tilt. For a few of the most deformed potentials that are still compatible with the Planck data we explicitly check that two-field corrections are negligible at times that the CMB modes cross the horizon. We keep $\alpha = 0.01$ fixed and vary $\log_{10}(\gamma) \in [-2.0,-1.5]$ and $\log_{10}(|\tilde \delta|) \in [-2,1]$. The results are shown in Fig.~\ref{fig:nsrnumerical} and ~\ref{fig:viableparameterspace}. Fig.~\ref{fig:nsrnumerical} shows the actual predictions for $n_s$ and $r$ and Fig.~\ref{fig:viableparameterspace} maps these to the implied constraints on $\tilde\delta$ and $\gamma$, given the $1\sigma$ and $2 \sigma$ confidence intervals of Planck. We confirm that the analytical results capture the predictions well if $\tilde\delta \ll1$, and that in this regime the inflationary observables are only mildly dependent on $\tilde \delta$. However, for larger deformations both $\tilde\delta$ and $\gamma$ become important to determine the observables, as one can see in both figures. On the other hand, changing $\alpha$ only shifts the value of $r$ up and down, and therefore the degeneracy between $c$ and $\alpha$ remains within the current CMB limits. However, 
since we require $c \geq 1$ the observational constraints demand $\alpha \lesssim 0.02$ which translates into the upper bound $r\lesssim 0.01$. Interestingly, even though we are on the boundary of small field and large field inflation, the tensor-to-scalar ratio approaches $r\sim 0.01$ already for deformations $\delta = \mathcal{O}(0.01)$. If in addition we only require minimal tuning on the parameters (that is, $|\tilde\delta|\sim 0.01$ and $\alpha\sim0.01$, and not smaller) the tensor-to-scalar ratio becomes bounded from below as well $r\gtrsim 10^{-4}$. As we see in Fig.~\ref{fig:viableparameterspace}, maintaining compatibility with observations and at the same time $c\geq 1$ forces us to keep $\delta = \frac{2\alpha}{(1-\alpha)\pi} \tilde \delta \lesssim  3 \times \frac{2\alpha}{(1-\alpha)\pi} $, maintaining a near-hybrid inflation regime. This need to tune $\delta$ small represents a requirement for an embedding of the inflationary mechanism into string theory, as there the quantities $c_{1}^\pm$ and $c_{2}^\pm$ will be determined by discrete data of the string compactification such as intersection numbers and p-form flux quanta. This will limit the tunability of $\delta$ and we will return to this question in section~\ref{sec:string}.
 
Let us finally stress that the inflationary predictions are very sensitive to the deformations of the hybrid potential, and moreover, we only considered the simplest modification so far. Therefore, the model lacks a clear prediction and it could be more informative to allude to a statistical approach as in~\cite{Dias:2018koa}.

 \subsubsection{Comparison with the effect of a non-zero instanton phase}
	\label{sec:Z2breakPhaseComp}

\begin{figure}[t!]
	\centering
	\includegraphics[scale=0.24]{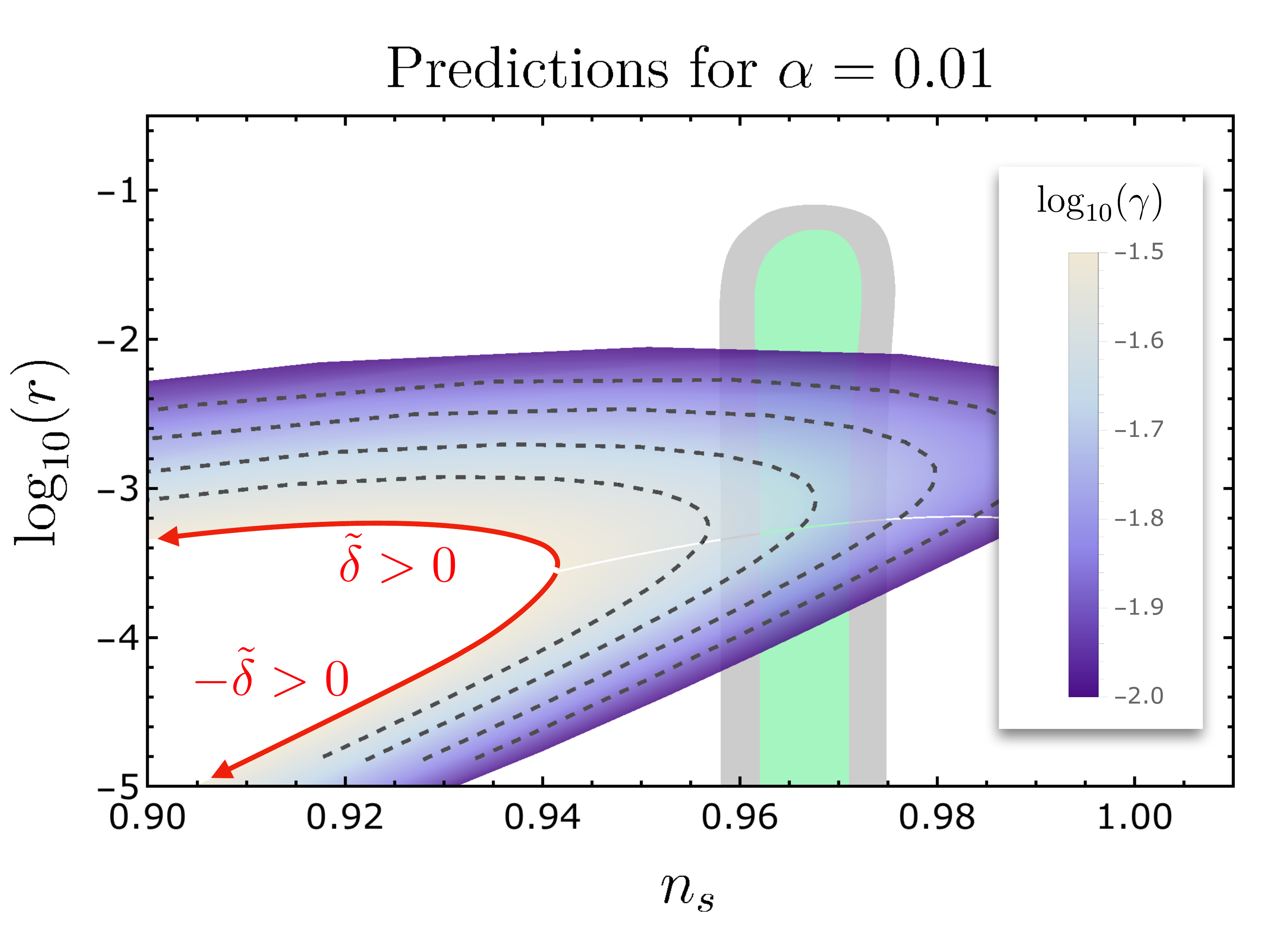}
	\caption{We compare the \textit{Planck} contours with the numerical predictions for $n_s$ and $r$ computed through the Hubble slow-roll parameters, where we vary $\log_{10}(|\tilde \delta|) \in [-2.0,1.0]$ and $\log_{10}(\gamma) \in [-2.0,-1.5]$, while fixing $\alpha =0.01$. The black dotted lines correspond to fixed $\gamma$ to $\{0.013, 0.017, 0.021, 0.025\}$, from right to left, respectively.  The region $\tilde\delta >0$ corresponds to inflationary trajectories with wide initial condition range of the type shown in Fig.~\ref{fig:ICSwithtrajectories} ending up in the false minimum at $\phi_1=2\pi/c_1$, $\phi_2=-{\cal O}(\tilde\delta)$.The complement  $\tilde\delta <0$ corresponds to trajectories which in Fig.~\ref{fig:ICSwithtrajectories} start with $\phi_2$-values beyond the saddle point with little allowed initial condition space, and end up in the true minimum at $\phi_1=\phi_2=0$.}
	\label{fig:nsrnumerical}
\end{figure} 

Let us now compare the results of section~\ref{sec:phase} and section~\ref{sec:Z2break}. By looking at eq.s~\eqref{etheta} and eq.~\eqref{edelta} we see that the phase $\vartheta$ coming from the instanton effects generating the potential and the asymmetry $\tilde\delta$ which is needed to break the $\mathbb{Z}_2$ vacuum degeneracy produce a similar correction to the first slow-roll parameter. Thus, by comparing the expressions for $\epsilon^{(\vartheta)}$ and $\epsilon^{(\delta)}$, we can extract a relation between these phases, which reads
	\begin{equation}\label{conversionthetadelta}
	\vartheta = - \frac{2 \alpha}{1 + \alpha}  \tilde\delta \,\, .
	\end{equation}
The above relation was derived analytically and it holds only for  $\vartheta \lesssim  \alpha$. This means that in this regime $\vartheta$ and $\tilde{\delta}$ are degenerate. The predictions for the tensor-to-scalar ratio and the spectral tilt in the presence of $\vartheta$ in this regime (Fig.~\ref{fig3} the small, central purple contour) and of $\tilde{\delta}$ (Fig.~\ref{fig:nsrgammadelta}) are the same once the values of $\vartheta$ and $\tilde\delta$ are related by the factor of eq.~\eqref{conversionthetadelta} (notice also that in Fig.~\ref{fig3} $\alpha$ is varied and $ c=1 $ is fixed while in~\ref{fig:nsrgammadelta} $\alpha$ is fixed while $\gamma\equiv \alpha c^2$ is varied). For $\vartheta > \alpha$ we could not perform the same analytical comparison, thus we cannot prove the degeneracy for all values of our parameters. The complete contours were obtained with the slow-roll approximation in Fig.~\ref{fig3} whereas for Fig.~\ref{fig:nsrnumerical} we solved the full equation of motion.

	\begin{figure}[t!]
	\centering
	\includegraphics[scale=0.6]{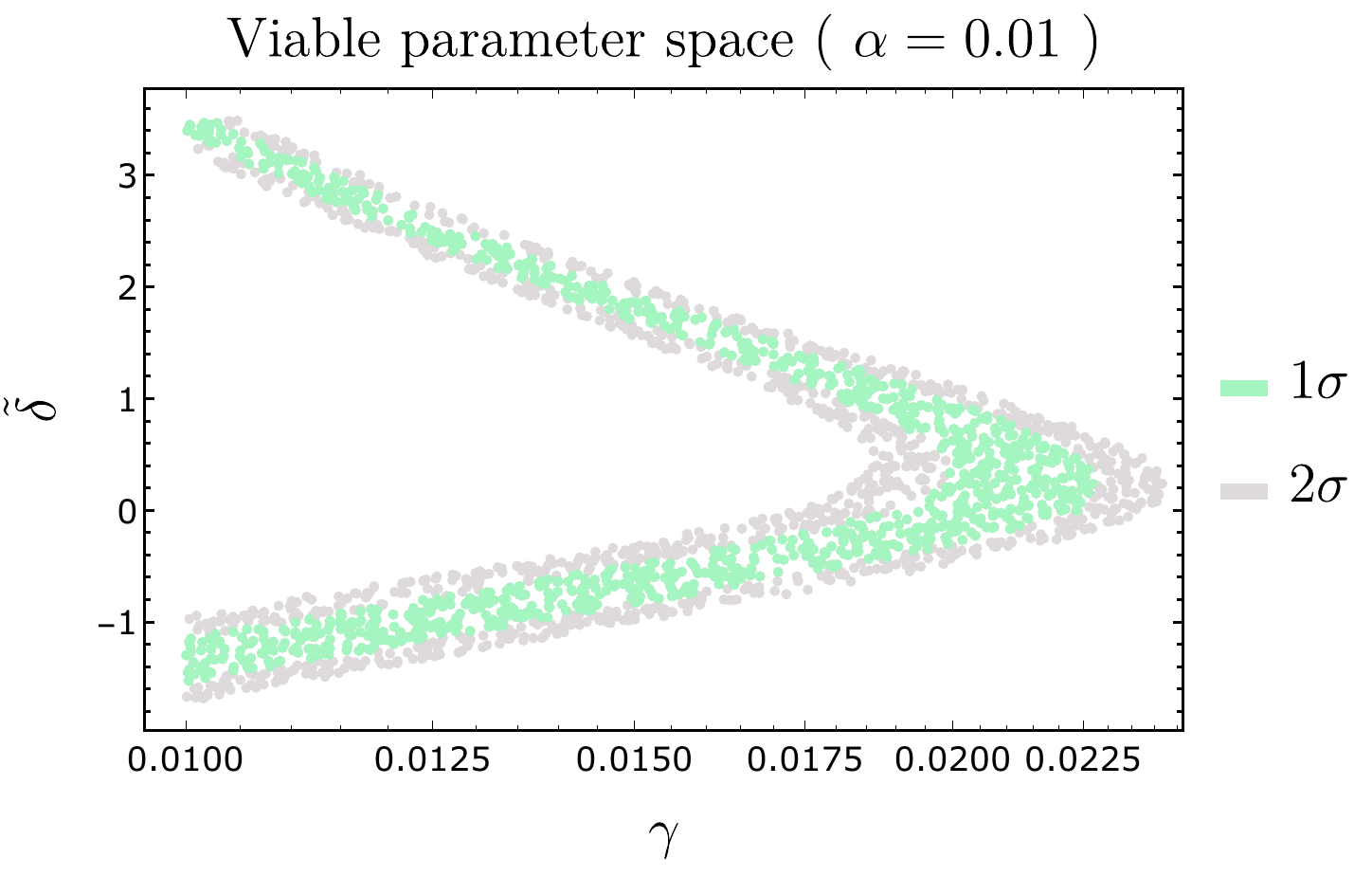}
	\caption{We randomly sample points in the same parameter range as Fig.~\ref{fig:nsrnumerical} (i.e. fixing $\alpha=0.01$) and select the ones within the $1\sigma$ (green) and $2\sigma$ (grey) \textit{Planck} confidence contours. The corresponding values for $\tilde\delta$ and $\gamma$ are shown. In the region $\tilde\delta >0$ the points corresponds to inflationary trajectories of the type shown in Fig.~\ref{fig:ICSwithtrajectories} ending up in the false minimum at $\phi_1=2\pi/c_1$, $\phi_2=-{\cal O}(\tilde\delta)$. In the complement  $\tilde\delta <0$ the points correspond to trajectories which in Fig.~\ref{fig:ICSwithtrajectories} start with $\phi_2$-values beyond the saddle point, and end up in the true minimum at $\phi_1=\phi_2=0$.}
	\label{fig:viableparameterspace}
\end{figure}

\subsection{Comments about eternal inflation and vacuum decay}\label{sec:tunnelEI}

We shortly describe two observations here, while we leave a more detailed analysis of these aspects for future work. The  inflationary valley of our harmonic hybrid model by necessity contains an inflationary saddle point with $\epsilon_V=0$, $|\eta_V|\ll 1$. We can now apply a comparison between the variance $\langle\delta\phi^2\rangle_q=H^2/(4\pi^2)$ of quantum fluctuations of the light inflaton scalar in near-dS space-time and the classical slow-rolling speed $\dot\phi = -V'/(3H)$ to argue for the presence of eternal inflation~\cite{Steinhardt:1982kg,Vilenkin:1983xq,Linde:1986fd} driven by quantum diffusion near the inflationary saddle point in our model. Arguments based on this comparison between quantum diffusion and classical rolling as reviewed e.g. in~\cite{Linde:2005ht,Guth:2007ng} show that the relevant criterion $\epsilon_V \lesssim V/(12\pi^2)$, $|\eta_V|<1$ when applied to exact hybrid inflation ($\delta=0$), produces a region satisfying the criterion around the saddle point, which is several hundred times wider than the average quantum fluctuation size $H/(2\pi)$. For the exact hybrid limit $\delta=0$ our models thus supports slow-roll eternal inflation.

However, avoidance of domain walls dictates $\delta \neq 0$ and moreover the scenarios for string theory realization of harmonic hybrid inflation we later discuss indicate that $|\delta|\gtrsim 0.01$. For such values of $\delta$ we observe that the region potentially supporting eternal inflation around the inflationary saddle point shrinks drastically. Its width reduces to a few times $H/(2\pi)$, rendering the existence of robust eternal inflation for these $\mathbb{Z}_2$-symmetry broken models doubtful. A much more detailed analysis is clearly necessary. If the outcome were the continued existence of this tension between model viability in the string theory context and ability to support eternal inflation, this would suggest two possible interpretations: we may either use it as evidence against slow-roll eternal inflation, or conversely, as an opportunity to predict the most likely size of $\delta$ once a well-defined measure for eternal inflation is found.

In presence of the $\mathbb{Z}_2$-symmetry breaking our setup contains a whole set of non-degenerate local minima with vacuum energy splitting of order $V_0\delta^2$ whose vacuum energy increases with increasing $\phi_1$-distance from the set of global minima at ($\phi_1=0$ , $\phi_2=2\pi n/c$). Between the local minima of this mini-landscape there will be tunnelling instanton transitions described by Coleman and de Luccia (CdL)~\cite{Coleman:1980aw} (for very recent work re-analyzing CdL tunnelling using the Hamiltonian-Wheeler-DeWitt approach, see~\cite{deAlwis:2019dkc}).

We next recall, that the inflationary trajectories starting from wide initial conditions always end up in one of the two false minima at ($\phi_1=2\pi/c_1$ , $\phi_2\simeq 0$ or $\phi_2\simeq 2\pi/c$). As we expect these slow-roll trajectories due to their large region of initial condition space to dominate the inflationary slow-roll dynamics, we need to tune the vacuum energy of the false minima at ($\phi_1=2\pi/c_1$ , $\phi_2\simeq 0$ or $\phi_2\simeq 2\pi/c$) to match the current-day vacuum energy $\sim 10^{-120}\mpl^4$. The adjacent true minima ($\phi_1=0$ , $\phi_2=2\pi n/c$) then are comparatively deep AdS vacua with vacuum energy of order $-V_0\delta^2$. Applying the CdL tunnelling description to these dS-AdS vacuum neighbours, we find that the so-called bounce action of the CdL instanton scales as $B\sim 1/V_{current\,c.c.}\sim 10^{120}$. Hence, tunnelling out of the false post-inflationary minimum into the nearest global AdS minimum is highly suppressed, with a life-time $\tau=\Gamma^{-1}\sim e^B\sim 10^{10^{120}}$ similar to the dS Poincar\'e recurrence time. Again, a more detailed study of the complete coupled system of CdL up and down transitions between the various dS vacua of the $\mathbb{Z}_2$-broken setup is left for future work.

\section{Towards a  String Theory Embedding}
	\label{sec:string}
	Axions in string theory arise as integrals of $p$-form gauge potentials over nontrivial cycles of the compactification manifold. Consider a type IIB string theory compactified to 4d on an orientifolded Calabi-Yau three-fold. We assume a choice of 3-form fluxes such that they stabilize the complex structure moduli at a high mass scale while generating an effectively constant superpotential $W_0$~\cite{Giddings:2001yu}.	In the 4d $\mathcal{N}=1$ low-energy theory, $O3/O7$ planes project the K\"ahler moduli space in even and odd subspaces with dimensions $h^{1,1}_+$ and $h^{1,1}_-$ respectively. This forces a rearrangement of the scalar degrees of freedom into $h^{1,1}_-$ axion multiplets $G^a= c^a - \tau b^a$, $a=1,\dots,h^{1,1}_-$, coming from the 2-forms $B_2$ and $C_2$, and $h^{1,1}_+$ complexified K\"ahler moduli
	\begin{equation}\label{complexifiedkahler}
	T_i= \frac{1}{2} k_{ijk}t^j t^k + i \int_{D_i} C_4+\frac{1}{4}e^{\phi} k^i_{ab}G^a\left(G-\bar{G}\right)^{b}
	\end{equation}
	where $i=1,\dots,h^{1,1}_+$, the $t^i$ are the 2-cycle volumes and the
	$k_{ijk}$ are the triple intersection numbers~\cite{Grimm:2004uq}. In section~\ref{sec::C4} we shall focus on orientifold projections such that $h^{1,1}_-=0$, that is we consider only the axions $\theta_i$ coming from the integral of the $C_4$ potential. After that, in section~\ref{sec::C2} we will also discuss a possible setup with $C_2$ axions as their presence could potentially avoid the domain-wall problem in an easy way.
	
	\subsection{$C_4$ Axions}
	\label{sec::C4}
	 When considering only the $C_4$ axions $\theta_i$, the expression eq.~\eqref{complexifiedkahler} simplifies to 
	\begin{equation}
	T_i= \frac{1}{2} k_{ijk}t^j t^k + i \int_{D_i} C_4= \tau_i+ i \theta_i
	\end{equation}
	where the scalar field $\tau_i$ is the volume of the 4-cycle divisor $D_i$. The volume of the Calabi-Yau can then be expressed in units of $l_s$ as
	\begin{equation}
	\mathcal{V}= \frac{1}{6}\int_X J\wedge J\wedge J= \frac{1}{6} k_{ijk} t^i t^j t^k= \frac{1}{3}\tau_i t^i.
	\end{equation}

	At the classical level and in the absence of branes, each $\theta_i$ enjoys a continuous shift symmetry. However the presence of the $O3/O7$ orientifold planes induces D3- and D7-brane charges. In order to cancel their tadpole, we must include D3-branes and D7-branes in the compactification setup. Such branes will break the continuous shift symmetry of $\theta_i$ into a discrete one by inducing non-perturbative corrections to the superpotential.  Harmonic hybrid inflation can then be obtained if we allow stacks of multiply-wrapped D7-branes on some of the 4-cycles. We will now discuss one mechanism to get the harmonic hybrid potential eq.~\eqref{finalpotential} from an LVS embedding.
	
	Start with assuming the volume $\mathcal{V}$ of a Calabi-Yau three-fold to be large compared to $l_s$ and to have the following Swiss-cheese like shape
	\begin{equation}
	\mathcal{V}\sim  \alpha \tau_b^{3/2}-\beta\tau_s^{3/2}-\gamma\tau_{s_1}^{3/2}-\delta\tau_{s_2}^{3/2},
	\end{equation}
	where $\alpha,\beta,\gamma,\delta$ are constants depending on the geometry of the manifold. For an explicit example of a Calabi-Yau with such volume see e.g.~\cite{Cicoli:2012vw}. Employing an LVS-type string compactification should allow us to use moduli stabilization to obtain a mass hierarchy among the axions. This is necessary to reproduce the dynamics of hybrid inflation models in general. However, we consider two additional terms compared to the classic LVS Swiss-cheese volume~\cite{Balasubramanian:2005zx} because $\theta_b$, the axionic partner of $\tau_b,$ is almost massless as it receives scalar potential contributions from $T_b$-dependent non-perturbative corrections which are highly suppressed by the compactification volume, i.e. $m_{\theta_b}\sim e^{-\mathcal{V}^{2/3}}\sim 0 $ (see~\cite{Conlon:2005ki} for a derivation of this statement). Moreover, when one stabilizes $\tau_s$, it can be shown that its axion $\theta_s$ gets stabilized as well, and in such a way that they gain approximately the same mass $m_{\theta_s}\sim m_{\tau_s}\sim m_{3/2}$. Therefore, $\theta_s$ is a rather heavy particle and it is frozen during inflation. We then infer that these two axions are not good candidates to reproduce our harmonic hybrid inflation. Thus, we should add two more blow-up moduli and arrange for their axions $\theta_{s_1}$, $\theta_{s_2}$ to be ultra-light but more massive than the axion corresponding to the base. The axionic mass hierarchy obtained from compactification should be $  m_{\theta_b} \ll m_{\theta_{s_2}}< m_{\theta_{s_1}} \lll m_{\theta_s} $.
	
	In general, the K\"ahler potential receives corrections both in string loop expansion and in $\alpha'$. In a typical LVS setup, one chooses to include the contribution of the leading $\alpha'^3$ correction (coming from the ten-dimensional $R^4$-term~\cite{Becker:2002nn}) and neglects string loops contributions. The K\"ahler potential is then
		\begin{equation}
		\begin{aligned}
			K&=-2\log\left(\mathcal{V}+\frac{\xi}{2}\right) +\delta K_{g_s}+\delta K_{\mathcal{O}(\alpha'^4)}=\\
			&=K_0 +\delta K_{g_s}+\delta K_{\alpha'},
	\end{aligned}
\label{KahlerLVS}	
	\end{equation}
	where $\xi$ is a function of $g_s$ given by
	\begin{equation}
	\xi=\dfrac{-\chi(CY_3)\zeta(3)}{2(2\pi)^3g_s^{3/2}}.
	\end{equation}
	and $\chi(CY_3)$ is the Euler characteristic of the Calabi-Yau $3$-fold.
	In ~\eqref{KahlerLVS} by $\delta K_{g_s}$ (resp. $\delta K_{\alpha'}$) we collectively mean all the string loop corrections~\cite{Berg:2005ja,Berg:2007wt,Cicoli:2007xp} to the K\"ahler potential (resp. all $\alpha'$ corrections).
	Considering only $K_0$ and the leading $\alpha'$ correction, the F-term scalar potential can be split into three terms, namely
		\begin{equation}
	V_F^{K_0}=V_{{np}_1}+V_{{np}_2}+V_{\alpha'}
	\end{equation}
	where 
	\begin{eqnarray}
	&&V_{{np}_1}=e^{K_0}K_0^{j\bar{k}}\,\partial_{T_j}W \partial_{\bar{T}_k}\bar{W}\, ,\\
	&&V_{{np}_2}=e^{K_0}K_0^{j\bar{k}}\left[\bar{W}\partial_{T_j}\!W\partial_{\bar{T}_k}\!K_0+h.c.\right]\, ,\\
	&&V_{\alpha'}\sim \frac{3\xi|W|^2}{16\mathcal{V}^3} \, .
	\end{eqnarray}
	We will now address the form of the superpotential $W$ in our model. We allow the tree-level superpotential $W_0$ to receive non perturbative corrections $W_{np}$ of the type
	\begin{equation}
	\begin{split}\label{supotential}
	W&=W_0+W_{np}=\\
	&=W_0+A_s e^{-a_s T_s}+A_{s_2} e^{-a_{s_2} T_{s_2}}\\&\quad+A_{s_1} e^{-a_{s_1}\left(n_{s_1}^1 T_{s_1}+n_{s_1}^2T_{s_2}\right)}\\&\quad+A_{s_2} e^{-a_{s_2}\left(n_{s_2}^1 T_{s_1}+n_{s_2}^2T_{s_2}\right)}.
	\end{split}
	\end{equation}
	These corrections can be explained as follows. Consider 4 stacks of D7-branes. One stack wraps the 4-cycle associated to K\"ahler modulus $T_s$, and another one wraps the 4-cycle associated to $T_{s_2}$. These two stacks give rise to the usual non-perturbative corrections to the superpotential from gaugino condensation (i.e the second line of eq.~\eqref{supotential}). Then the other two stacks wrap two different cycles which are representative of two different divisor classes, each one being a linear combination of the divisor classes of $T_{s_1}$ and $T_{s_2}$. Finally, we allow for the D7-branes of these last stacks to wrap the cycle multiple times~\cite{Long:2014dta}. This information is encoded in the winding numbers $n_{s_i}^j$ and their inclusion modifies the superpotential corrections as in eq.~\eqref{supotential}. 
	
	Here we will anticipate the crucial point of this type of embedding:  in order to recover the effective potential eq.~\eqref{finalpotential} we will need to require $n_{s_2}^2<0$. From now on the negative sign will be extracted. We might worry about this linear combination with negative coefficients, for the following reason: in order for the non-perturbative corrections to the superpotential to arise, the stacks of branes must wrap rigid and ample divisors~\cite{Witten:1996bn, Bobkov:2010rf}. It is therefore natural to wonder if  the fact that $n^2_{s_2}<0$ necessarily spoils the ampleness condition. We argue that this is not always the case. In fact, the divisors corresponding to the K\"ahler moduli in the exponents could be themselves linear combinations of toric divisors, and not just toric divisors. Therefore we may be able to change base and re-write the exponents of eq.~\eqref{supotential} in a way in which only toric divisors appear, and with positive coefficients. If this is the case, we may be able to satisfy the requirements of rigidity and ampleness. A concrete example of this can be found in~\cite{Blumenhagen:2007sm}. We should now point out that the Calabi-Yau required for our model is certainly not present in the catalogue of Swiss-cheese solutions found in~\cite{Altman:2017vzk}, where only stack of branes wrapping toric divisors are considered. It would be very interesting to extend the catalogue of~\cite{Altman:2017vzk} including the possibility of wrapping branes on linear combinations of toric divisors, in order to find examples of Calabi-Yaus that can support this model as well as to widen the possibilities to realize LVS moduli stabilization.
	
Moduli stabilization with LVS mechanism forces
	\begin{equation}
		\partial_{T_s}W_{np}\sim \frac{W_0}{\mathcal{V}}.
	\end{equation}
	To achieve scale separation between the axions $\theta_{s_1}$ and $\theta_{s_2}$ with respect to $\theta_s$, we choose $a_{s_1}, a_{s_2}, \tau_{s_1}$ and $\tau_{s_2}$ such that
	\begin{equation}
	\partial_{T_{s_1}}W_{np}\sim\partial_{T_{s_2}}W_{np}\sim \epsilon\,\frac{W_0}{\mathcal{V}}\, ,
	\end{equation}
 where $\epsilon<1$. 
 If $\mathcal{V}^{-1}\ll\epsilon\ll 1$, then axions get stabilized after K\"ahler moduli. The dominant terms for the $\theta_{s_1}$, $\theta_{s_2}$ axions are then the ones of order $\mathcal{O}\left(\epsilon\, \mathcal{V}^{-3}\right)$. It can be shown that these are given by the $T_{s_i}\bar{T}_{s_j}(+ h.c.)$, $i,j=1, 2$, terms in $V_{np_2}$, namely
 \begin{equation}
 	V_{np_2}\left(\theta_{s_1}, \theta_{s_2}\right)=\frac{1}{\mathcal{V}^2}\left[K_0^{i\bar{j}}W_0\partial_{T_{s_i}}\!\!W_{np}\partial_{\bar{T}_{s_j}}\!\!K_0+h.c.\right].
 \end{equation}
 Plugging in this equation the derivatives of the superpotential
 \begin{eqnarray}
 	&&\begin{split}\partial_{T_{s_1}}W_{np}= &- a_{s_1} A_{s_1} n_{s_1}^1 e^{-a_{s_1}\left(n_{s_1}^1 T_{s_1}+n_{s_1}^2T_{s_2}\right)}\\&-a_{s_2} A_{s_2}n_{s_2}^1 e^{-a_{s_2}\left(n_{s_2}^1 T_{s_1}-n_{s_2}^2T_{s_2}\right)}\end{split}\\
 	&&\begin{split}\partial_{T_{s_2}}W_{np}=&-a_{s_2} A_{s_2}e^{-a_{s_2}T_{s_2}}\\&- a_{s_1} A_{s_1} n_{s_1}^2 e^{-a_{s_1}\left(n_{s_1}^1 T_{s_1}+n_{s_1}^2T_{s_2}\right)}\\& -a_{s_2} A_{s_2}n_{s_2}^2 e^{-a_{s_2}\left(n_{s_2}^1 T_{s_1}-n_{s_2}^2T_{s_2}\right)}\end{split}
 \end{eqnarray}
	and using the fact that $K_0^{i\bar{j}}K^0_i = - \frac{1}{4}\tau_j$, the potential for the $\theta_{s_1}, \theta_{s_2}$ axions can be written as
	\begin{equation}
		\begin{split}
		&V_{np_2}\left(\theta_{s_1}, \theta_{s_2}\right)=
		\frac{2W_0}{\mathcal{V}^2}\tau_{s_2} a_{s_2} A_{s_2}e^{-a_{s_2}\tau_{s_2}}\cos\left(a_{s_2}\theta_{s_2}\right)\\
		&+\frac{2W_0}{\mathcal{V}^2}a_{s_1} A_{s_1}\!\left(n_{s_1}^1\tau_{s_1}+n_{s_1}^2\tau_{s_2}\right)e^{-a_{s_1}\!\left(n_{s_1}^1\tau_{s_1}+n_{s_1}^2\tau_{s_2}\right)}\cdot\\&\quad\quad\cdot\cos\left[a_{s_1}\!\left(n_{s_1}^1\theta_{s_1}+n_{s_1}^2\theta_{s_2}\right)\right]\\
		&+\frac{2W_0}{\mathcal{V}^2}a_{s_2} A_{s_2}\!\left(n_{s_2}^1\tau_{s_1}-n_{s_2}^2\tau_{s_2}\right)e^{-a_{s_2}\!\left(n_{s_2}^1\tau_{s_1}-n_{s_2}^2\tau_{s_2}\right)}\cdot\\&\quad\quad\cdot\cos\left[a_{s_2}\!\left(n_{s_2}^1\theta_{s_1}-n_{s_2}^2\theta_{s_2}\right)\right].
		\end{split}
	\end{equation}

	Our harmonic hybrid inflation potential eq.~\eqref{finalpotential} is recovered if we take $\theta_{s_i} = \phi_i/f_{s_i}$ as explained below,
	\begin{equation}
		\tilde{\Lambda}^4=\frac{2W_0}{\mathcal{V}^2}a_{s_2}\tau_{s_2} A_{s_2}e^{-a_{s_2}\tau_{s_2}}
	\end{equation}
	and by requiring the prefactors of the two mixed cosines to be equal:
	\begin{equation}
	\begin{split}
	\Lambda^4 &= \frac{2W_0}{\mathcal{V}^2}a_{s_1} A_{s_1}\!\left(n_{s_1}^1\tau_{s_1}+n_{s_1}^2\tau_{s_2}\right)e^{-a_{s_1}\!\left(n_{s_1}^1\tau_{s_1}+n_{s_1}^2\tau_{s_2}\right)}=\\  &= \frac{2W_0}{\mathcal{V}^2}a_{s_2} A_{s_2}\!\left(n_{s_2}^1\tau_{s_1}-n_{s_2}^2\tau_{s_2}\right)e^{-a_{s_2}\!\left(n_{s_2}^1\tau_{s_1}-n_{s_2}^2\tau_{s_2}\right)} \quad .
		\end{split}
	\end{equation}

	We now discuss how to incorporate the $\mathbb{Z}_2$-symmetry breaking. For this, we need to keep in mind that we need stacks of at least 2 coincident D7-branes for each instanton effect to generate  a superpotential and in turn the scalar potential above. Hence we have that $a_{s_i}=2\pi/N_{s_i}$ with $N_{s_i}\geq 2$ in the cosine terms of the scalar potential. The effective coefficients of the two axion fields in the two cosine terms in the scalar potential thus read
	\begin{equation}
		\begin{split}
		c_{1}^+ \equiv &\, \frac{2\pi}{f_{s_1}} \frac{n^1_{s_1}}{N_{s_1}}, \qquad c_{1}^- \equiv  \,\frac{2\pi}{f_{s_1}} \frac{n^1_{s_2}}{N_{s_2}}, \\
		c_{2}^+ \equiv & \,\frac{2\pi}{f_{s_2}} \frac{n^2_{s_1}}{N_{s_1}}, \qquad c_{2}^- \equiv  \,\frac{2\pi}{f_{s_2}} \frac{n^2_{s_2}}{N_{s_2}},
		\end{split}
		\end{equation}
	where for simplicity we assume that the string compactification has produced diagonal kinetic terms for the two axions of the form ${\cal L}_{kin}=\frac12 f_{s_1}^2(\partial\theta_{s_1})^2+\frac12 f_{s_2}^2(\partial\theta_{s_2})^2$, implying canonically normalized axion fields $\theta_{s_i} = \phi_i/f_{s_i}$.
	Hence, maximizing the field range of the inflaton $\phi_2$ and thus $r$ while keeping $f_{s_2}\leq M_{\rm P}$ sub-Planckian implies a relation between wrapping numbers and gauge group ranks: $n^2_{s_i}=N_{s_i}/2\pi$. The simplest choice achieving this would be $N_{s_1}=N_{s_2}=6$ which produces $c_{2}^\pm\simeq 1$. 
	
	Next, maintaining inflationary dynamics behaving nearly like exact hybrid inflation requires a mass hierarchy between $m_{\phi_1}\gg H\gg m_{\phi_2}$ and a small amount of $\mathbb{Z}_2$-symmetry breaking by setting $c_{1}^\pm=c_{1}\pm\delta_c$ with $\delta_c\ll c_{1}$ as in section~\ref{sec:Z2break}. The first condition requires $c_{1}^\pm\gtrsim 10$ implying $n^1_{s_i}\gtrsim 10\, n^2_{s_i}$. Since the $n^2_{s_i}$ are integers, their smallest difference $\delta_c=1$ implies a 10\%-level $\mathbb{Z}_2$-symmetry breaking.  A level of 1\% $\mathbb{Z}_2$-breaking is possible for the same $\delta_c$ for a choice of $N_{s_1}=N_{s_2}=60$ and wrapping numbers $n^1_{s_1}=100+\delta_c$, $n^1_{s_2}=100-\delta_c$ and $n^1_{s_i}\approx10n^2_{s_i}$, for $n^2_{s_i}=10$, while maintaining the $\phi_2$ field range and $\phi_1$-$\phi_2$ mass hierarchy.
	
	\subsection{$C_2$ Axions}
	\label{sec::C2}
	We now discuss a second scenario for generating the potential eq.~\eqref{finalpotential}. $C_2$ axions present in Calabi-Yau orientifolds with $h^{1,1}_->0$ can acquire a periodic scalar potential if the relevant stack of D7-branes present in the compactification is magnetized \cite{Long:2014dta}. Then, the gauge kinetic function is modified and depends holomorphically also on the $h^{1,1}_-$ axion multiplets $G^a$. In this way, the continuous shift symmetry of the odd axion $c^a$ is broken to a discrete one at the level of the superpotential $W$ when we consider also its non-perturbative corrections. The other axions $\theta_i$ can be stabilized at a higher scale than $c^a$ by considering one additional unmagnetized stack wrapping another representative of the homology class. With this setup, gaugino-condensation on
		different D7-branes stacks gives rise to the non-perturbative corrections of the form\
		\begin{equation}
		W_{np}= \sum_{\zeta} A_\zeta e^{-a_\zeta \left(T_\zeta+i k_{\zeta mn}F^{m}_\zeta\left(G^n+\frac{\tau}{2}F_\zeta^{n}\right)\right)}\quad.
		\end{equation}
		Here   
		\begin{itemize}
			\item $k_{mn}^\zeta$ are the triple intersection numbers between the divisors $\zeta=D_1\ldots D_3$, and the odd 2-cycles $\Sigma_{m/n}^{(-)}$.
			\item $m,n$ run from 1 to $h^{1,1}_-$ and $\zeta$ runs over the wrapped divisors $D_1 \ldots D_{h^{1,1}_+}$.
			\item $F_{D_i}^{m}$ are the fluxes each stack on the divisor $D_i$ carries, but they also refer to only one ($m$) of the two $C_2$ axions each time.
		\end{itemize}
		
		In this setup we shall consider $h^{1,1}_-=2$ and $h^{1,1}_+=3$.
		As before, consider the complex structure moduli to be stabilized at a higher scale. Assume also that the $\tau_{D_i}$, $\theta_{D_i}$ and $b^m$ are stabilized by a combination of the LVS mechanism, string-loop corrections and/or higher order F-term contributions, D-terms and by unmagnetized D-branes, and that the $b^a=0$ are at their minimum. Next, assume that both odd cycles $\Sigma_{1,2}^{(-)}$ intersect with the divisors corresponding to $\tau_{D_1}$ and $\tau_{D_2}$ while only  $\Sigma_{2}^{(-)}$ intersects with the divisor corresponding to $\tau_{D_3}$. These are conditions on the non-vanishing intersection numbers. They combine with the splitting of the intersection numbers indices $k_{ABC}$, $k=1\ldots 5$ into the even indices $A=(\zeta,m)$ and the restructuring of the 2-cycle volume moduli into 4-cycle ${\cal N}=1$ K\"ahler moduli containing 4-cycle volumes $\tau_\zeta$ and the 2-form axion chiral multiplets $G^m$. If the inversion relation between the 2-cycles $v^A$ and the $(\tau_\zeta,G^m)$ can be performed explicitly, this may result -- to provide a simple example -- in a volume schematically of the form
		\begin{equation}
		\begin{split}
		\mathcal{V}\sim & \left(\tau_{D_1}+k^{D_1}_{mn}b^mb^n\right)^{3/2}-\left(\tau_{D_2}+k^{D_2}_{mn} b^mb^n\right)^{3/2}\\
		& -\left(\tau_{D_3}+k^{D_3}_{22}(b^2)^2\right)^{3/2}\quad,\quad \mbox{where}:\;\;m,n=1,2\quad.
		\end{split}
		\end{equation}

	In this case, the potential for the $c^m$ axions takes the form
	\begin{equation}
	\begin{split}
	V&=\sum_{\zeta} \Lambda_\zeta^4 \left[ 1-\cos\left(a_\zeta k^\zeta_{mn} F_\zeta^m c^n\right)\right]\\
	& =\Lambda_{D_1}^4\left[1-\cos\left( \left(a_{D_1} k_{11}^{D_1} F^1_{D_1}+a_{D_1} k_{21}^{D_1} F^2_{D_1}\right)c^1\right.\right.\\
	&\left.\left.\qquad\qquad \qquad+\left(  a_{D_1} k_{12}^{D_1} F^1_{D_1}+a_{D_1} k_{22}^{D_1} F_{D_1}^2\right) c^2\right)\right]+\\
	&\quad\, \Lambda_{D_2}^4\left[1-\cos\left( \left(a_{D_2} k_{11}^{D_2} F^1_{D_2}+a_{D_2} k_{21}^{D_2} F_{D_2}^2\right)c^1\right.\right.\\
	&\left.\left.\qquad\qquad \qquad+\left(  a_{D_2} k_{12}^{D_2} F^1_{D_2}+a_{D_2} k_{22}^{D_2} F_{D_2}^2\right) c^2\right)\right]+\\
	&\quad\, \Lambda_{D_3}^4\left[1-\cos\left( a_{D_3} k_{22}^{D_3} F_{D_3}^2 c^2\right)\right].
	\end{split}	
	\end{equation}

	Next, the fact that the stacks of $N_{D_i}$ D7-branes are wrapped on each divisor $D_i$ producing the instanton contributions in $W$ implies that $a_{D_i}=2\pi/N_{D_i}$. The intersection numbers $k^\zeta_{mn}$ are fixed numerical quantities for a given CY manifold, while the fluxes $F_{D_i}^m$ are integers. Finally, canonically normalizing the axions $c^{m}$ will replace $c_1,c_2$ with $\phi_1/f_1,\phi_2/f_2$, where $f_1,f_2$ represent the axion decay constants of the canonically normalized axion fields $\phi_1$, $\phi_2$. Hence, the resulting terms
	\begin{equation}\label{eq:c2Embed}
	\begin{split}
	\left(a_{D_1} k_{11}^{D_1} F^1_{D_1}+a_{D_1} k_{21}^{D_1} F^2_{D_1}\right)\frac{\phi_1}{f_1}\equiv &\, c_{1}^+\phi_1\\
	\left(a_{D_2} k_{11}^{D_2} F^1_{D_2}+a_{D_2} k_{21}^{D_2} F_{D_2}^2\right)\frac{\phi_1}{f_1}\equiv & \,c_{1}^-\phi_1\\
	\left(a_{D_1} k_{12}^{D_1} F^1_{D_1}+a_{D_1} k_{22}^{D_1} F_{D_1}^2\right)\frac{\phi_2}{f_2}\equiv &\, c_{2}^+\phi_2\\
	\left(a_{D_2} k_{12}^{D_2} F^1_{D_2}+a_{D_2} k_{22}^{D_2} F_{D_2}^2\right)\frac{\phi_2}{f_2}\equiv &\, c_{2}^-\phi_2
	\end{split}
	\end{equation}
	will typically have $c_{1}^+\neq c_{1}^-$  and $c_{2}^+\neq c_{2}^-$ (unless particular combinations of the $a_{D_{1,2}}$, the triple intersection numbers, and the gauge flux quanta appearing in eq.~\eqref{eq:c2Embed} are chosen in very particular combinations). At most, they typically can be tuned by choosing the flux integers to be nearly the same to some finite accuracy, leading automatically to the situation of section~\ref{sec:Z2break}. Hence, embedding harmonic hybrid inflation into string theory along these lines \emph{generically} avoids the hybrid inflation domain wall problem. Thus, by tuning appropriately the triple intersection numbers and the flux quanta, the potential eq.~\eqref{fullpotential} can be recovered.
	
	The perks of using the $C_2$ axions instead of $C_4$ axions are that one has more freedom in tuning parameters to recast the needed potential, and that interestingly they are better for phenomenological purposes as well.

\subsection{Outlook - thraxions}

	So far we have discussed two possibilities to obtain harmonic hybrid inflation potentials using non-perturbative quantum effects to provide the required scale suppression and periodicities. However, even in the absence of perturbatively exactly massless harmonic 2-form axions, recent work pointed out that CY compactifications with warped multi-throat regions  
	contain ultra-light radial KK-modes of the would-be $C_2$-axion harmonic zero mode.
These so-called `thraxions' acquire a periodic cosine potential due to the compactness of the CY manifold with periodicity enhanced by flux monodromy to roughly Planckian values~\cite{Hebecker:2018yxs}. 

Furthermore, a recently found database listing all ambient-space descending $O3/O7$-orientifolds of the set of Complete Intersection Calabi-Yaus (CICYs) contains thousands of explicit examples of compact CYs whose topology supports the existence of such thraxions~\cite{Carta:2020ohw}. For other $O3/O7$-orientifold constructions in the Kreuzer-Skarke database see~\cite{Altman2017ra,Gao:2013pra}.

As each pair of homologically related conifolds generate one thraxion with its scalar potential, a compactification with multiple double-throats may lead to a sum-separable scalar potential of multiple cosine terms for multiple thraxions. Since the thraxion kinetic terms will generically have mixing due to the structure of the intersection numbers of the underlying CY manifold, the scalar potential after diagonalization of the kinetic term may in certain cases very well take the form hereby proposed for harmonic hybrid inflation. We leave the study of this intriguing possibility for future work.
	
	\section{Conclusions}
	Let us now discuss what we have found. We have seen that a sector of two axion fields with a purely non-perturbatively generated scalar potential contains a mechanism for realizing hybrid inflation. The structure of scalar potential turns out to be highly constrained by the discrete shift symmetries of the axions. Bottom-up, we managed to construct the most simple toy model field theory setup consisting of three cosine potential terms for two axions producing the hybrid inflation coupling structure. Next, we analysed the influence of initial condition choices on the occurrence of successful inflation, showing that harmonic hybrid inflation generates observationally viable slow-roll inflation for a wide range of initial conditions. 
	
	Hybrid inflation setups generating the hybrid behaviour by means of a bi-quadratic coupling between the inflaton and the waterfall field usually show a $\mathbb{Z}_2$-symmetry between the two degenerate final vacua of the scalar field trajectory. This $\mathbb{Z}_2$ symmetry leads to domain wall production overclosing the late universe. As a remedy, we show that controlled $\mathbb{Z}_2$-symmetry breaking of adjacent axion vacua is achievable for the price of a moderate amount of tuning of the coefficient of the axions appearing in the arguments of the cosine terms in the scalar potential. Including this $\mathbb{Z}_2$-symmetry breaking when otherwise minimally tuning the setup leads to prediction of primordial tensor modes with the tensor-to-scalar ratio in the range $10^{-4}\lesssim r \lesssim 0.01$, directly accessible to upcoming CMB observations.
	
	Moreover, we found that the presence of an inflationary saddle point with $\epsilon=0$ and $|\eta|\ll 1$ produces a regime of quantum diffusion driven slow-roll eternal inflation as long as the $\mathbb{Z}_2$-symmetry breaking parameter $\delta\ll1$ is small enough. As we need $\delta >0$ by some finite amount is necessary to avoid the domain wall problem, this may indicate that harmonic hybrid inflation shrinks the field space region available for eternal inflation and might conceivably even favor its absence.~\footnote{We leave possible connections between this observations and attempts to connect a better understanding of potential constraints on de Sitter space with the absence of slow-roll eternal inflation in the context of the swampland conjectures for future work.} We observe furthermore, that harmonic hybrid inflation looks compatible with certain UV arguments supporting constraints $f\lesssim \mpl$ and $\Delta\phi_{60}\lesssim \mpl$ on the axion periodicity and slow-roll field range, respectively. 
	
	However, we need to be more cautious. Note that the fine-tuning of initial conditions needed to generate enough inflation quickly increases once we choose the larger of the two axion decay constants sub-Planckian by a significant amount. As this implies rather $f_{\phi_2}\lesssim \mpl$ instead of $f_{\phi_2}\ll\mpl$, a concrete realization of our mechanism needs to check if higher instanton harmonics with instanton actions of e.g. gravitational instantons generically scaling as $S_{inst.,n}\sim n \mpl/f_{\phi_2}$ remain sufficiently suppressed (see e.g.~\cite{Banks:2003sx,Kaloper:2008fb,Hebecker:2018ofv}). We leave the necessary more detailed analysis, clearly dependent on UV input, for future more complete string theory realizations.

	Finally we also outlined several roads towards constructing harmonic hybrid inflation in type IIB string theory on Calabi-Yau orientifold compactifications. We provide arguments that the relevant scalar potential can arise either from $C_4$ axions via multiple-wrapped D7-brane stacks, from $C_2$-axions acquiring scalar potential from D7-brane stacks magnetized by quantized gauge flux, or potentially from thraxions in bifid-throat regions of CY orientifolds.

	\bigskip
	
	\centerline{\bf \large Acknowledgments}
	
	We thank J. Moritz and F. Muia for helpful discussions. AW would like to thank N. Kaloper for several very elucidating conversations. NR is supported by the Deutsche Forschungsgemeinschaft under Germany's Excellence Strategy - EXC 2121 ``Quantum Universe'' - 390833306. FC, AW and YW are supported by the ERC Consolidator Grant STRINGFLATION under the HORIZON 2020 grant agreement no. 647995.

\appendix
	
\section{Critical points and some number theory}\label{app:criticalpoints}

In order to solve the two-field dynamics of harmonic hybrid inflation numerically, it is useful to know the location of the saddle points and minima to set appropriate initial conditions.  For our toy model in eq.~\eqref{potential} the fundamental domain (centralized around the minimum) of the fields is given by $c_1 \phi_1 \in [0, 2\pi)$ and $c_2 \phi_2 \in [0, 2\pi)$. It's easy to check that some interesting points are located at 
\begin{subequations}
\begin{align}
(c_1 \phi_1, c_2 \phi_2) & = (0,0) \quad  && \text{(global minimum)}, \\
(c_1 \phi_1, c_2 \phi_2) & = (0,\pi) \quad  && \text{(global maximum)}, \\
(c_1 \phi_1, c_2 \phi_2) & = (\pi,\tfrac{\pi}{2})  \quad && \text{(waterfall transition)}, \\
(c_1 \phi_1, c_2 \phi_2) & = (\pi,\pi)  \quad && \text{(inflation saddle point)},
\end{align}
\end{subequations}
assuming that $\alpha>1$. For the assumptions listed in section \ref{sec:singlefield} the region around the saddle point $(\pi,\pi)$ is suitable to start an extended period of inflation.
Consider, however, a more general potential of the form 
\begin{equation}
\begin{split}
V=\,&\Lambda^4_1[1- \cos\left({d}_{1}\phi_1+{c}_{1}\phi_2\right)]\\&+ \Lambda^4_2[1-\cos\left({d}_{2}\phi_1-{c}_{2}\phi_2\right)]\\&+ \Lambda^4_3[1-\cos\left({c}_{3}\phi_2\right)]\ .
\end{split}
\end{equation}	
In this case it becomes non-trivial to find a set of equivalent points. First of all, assuming that $c_i$ and $d_j$ are integers, the fundamental domain is now given by $\text{gcd}(d_1,d_2) \phi_1 \in [0, 2\pi)$ and $\text{gcd}(c_1,c_2,c_3) \phi_2 \in [0, 2\pi)$ which may now contain multiple minima and suitable saddle points to host inflation. We might not be able to find all critical points analytically, however, we may still find a non-trivial subset of them by looking for field coordinates where each cosine is equal to $+1$ or $-1$ (such that the gradient of the potential vanishes). If all cosines equal $+1$ ($-1$) we are clearly in the global minimum (global maximum) of the potential. For the other combinations we have to look at the Hessian, whose determinant is given by
\begin{equation}
\begin{split}
\det \nabla \nabla V  = &\ \left( \sum_{i=1}^2 \Lambda_i^4 d_i^2 \sigma_i \right)\left( \sum_{j=1}^3 \Lambda_j^4 c_j^2 \sigma_j \right)\\
&\ -\left(  \Lambda_1^4 c_1 d_1 \sigma_1-\Lambda_2^4 c_2 d_2 \sigma_2\right)^2 \ .
\end{split}
\end{equation}
Here $\sigma_i  \in \{+1,-1\} $ is the value that each cosine takes. We can rewrite this as 
\begin{equation}
\begin{split}
\det \nabla \nabla V  = &\  \Lambda_1^4 \Lambda^4_2 (d_1 c_2 +c_1 d_2)^2 \sigma_1 \sigma_2  \\
 + &\ \Lambda_1^4 \Lambda^4_3 d_1^2 c_3^2 \sigma_1 \sigma_3 \\
+ &\ \Lambda_2^4 \Lambda^4_3 d_2^2 c_3^2 \sigma_2 \sigma_3\ . \\
\end{split}
\end{equation} 
If, for instance, similarly to the toy model, $\Lambda^4_3  c_3^2\left(\Lambda_1^4 d_1^2+\Lambda_2^4 d_2^2\right) > \Lambda_1^4 \Lambda^4_2 (d_1 c_2 +c_1 d_2)^2$, we have a saddle point for $\sigma_1 = \sigma_2 = -\sigma_3 =  1$. More generally, to inspect any of these critical points, if they exist, we have to solve the following set of equations 
\begin{subequations}
\begin{align}
{c}_{1}\phi_2+{d}_{1}\phi_1 = &\ m_1 \pi+  2\pi N_1, \\
{c}_{2}\phi_2 -{d}_{2}\phi_1= &\ m_2 \pi +  2\pi N_2, \\
{c}_{3}\phi_2 = &\ m_3 \pi +  2\pi N_3\ ,
\end{align}
\end{subequations}
with $m_i \in \{0, \pm 1\} $.
Eliminating $\phi_1$ and $\phi_2$ we find the following constraint 
\begin{equation}
c_3d_2N_1+c_3d_1 N_2 -\gamma N_3 + \frac{c_3\mu - \gamma m_3 }{2} = 0,
\end{equation}
where we defined $\gamma \equiv d_2 c_1 + d_1 c_2$ and $\mu \equiv d_2 m_1+d_1 m_2$. Now, if 
\begin{equation}
r \equiv \frac{c_3\mu - \gamma m_3 }{2 \text{gcd}(c_3d_2, c_3d_1, \gamma )} \in \mathbb{Z}
\end{equation}
a solution for $N_i$ is given by
\begin{subequations}
\begin{align}
N_1 = &\ -r n_1, \\
N_2 = &\ -r n_2, \\
N_3 = &\  r n_3,
\end{align}
\end{subequations}
where $n_i$ solves the diophantine equation
\begin{equation}
c_3d_2n_1+c_3d_1 n_2 + \gamma n_3 = \text{gcd}(c_3d_2, c_3d_1, \gamma )\ .
\end{equation}
A solution for this equation can be found by applying the Euclidean algorithm. The corresponding critical point is then located at 
\begin{subequations}
\begin{align}
\phi_2 = &\ \frac{m_3 +2 N_3}{c_3} \pi, \\
\phi_1 = &\  \frac{m_1 +2 N_1}{d_1} \pi -c_1 \phi_2 \\
 =&\ c_2 \phi_2 -\frac{m_2 +2 N_2}{d_2} \pi\ . \nonumber
\end{align}
\end{subequations}
Where we should map the particular solution back to the fundamental domain. \\

Let us highlight three examples where we go through the steps outlined above and see which critical points we find.
\begin{itemize}
\item The hybrid model with $c_i = 1$ and $d_i =10$, but with the amplitudes arbitrary. We find the following critical points:
\begin{subequations}
\begin{align}
(\phi_1, \phi_2) & = (0,0) \quad  && V = 0, \\
(\phi_1, \phi_2) & = (0,\tfrac{\pi}{10}) \quad  &&  V = 2(\Lambda_1^4+\Lambda_2^4), \\
(\phi_1, \phi_2) & = (\pi,\tfrac{\pi}{10})  \quad &&  V = 2\Lambda_3^4, \\
(\phi_1, \phi_2) & = (\pi,0)  \quad &&  V = 2(\Lambda_1^4+\Lambda_2^4+\Lambda_3^4) ,
\end{align}
\end{subequations}
where the first and last points are clearly the global minimum and maximum. The two middle ones are saddle points if the determinant of the Hessian is negative, i.e. if $50(2 \Lambda_1^4 \Lambda_2^4 - \Lambda_1^4 \Lambda_3^4  - \Lambda_2^4 \Lambda_3^4) < 0$. 
\item The slightly deformed hybrid model with $c_i = 1$ and $d_1 =9$ and $d_2 =11$. We find:
\begin{subequations}
\begin{align}
(\phi_1, \phi_2) & = (0,0) \quad  && V = 0, \\
(\phi_1, \phi_2) & = (0,\pi) \quad  &&  V = 2(\Lambda_1^4+\Lambda_2^4), \\
(\phi_1, \phi_2) & = (\pi,\pi)  \quad &&  V = 2\Lambda_3^4, \\
(\phi_1, \phi_2) & = (\pi,0)  \quad &&  V = 2(\Lambda_1^4+\Lambda_2^4+\Lambda_3^4) .
\end{align}
\end{subequations}
Now the condition of the two middle critical point to be saddle points reads $100 \Lambda_1^4 \Lambda_2^4 - \tfrac{121}{2} \Lambda_1^4 \Lambda_3^4  - \tfrac{81}{2} \Lambda_2^4 \Lambda_3^4 < 0$. 
\item The much deformed hybrid model with $\vec{c} = \{3,6,15\}$ and $\vec{d} = \{40,45\}$. We find six critical points:
\begin{subequations}
\begin{align}
(\phi_1, \phi_2) & = (0,0) \quad  && V = 0, \\
(\phi_1, \phi_2) & = (0,\tfrac{\pi}{5}) \quad  &&  V = \Lambda_2^4, \\
(\phi_1, \phi_2) & = (\tfrac{\pi}{15},\tfrac{8\pi}{25})  \quad &&  V = 2\Lambda_3^4+\Lambda_1^4, \\
(\phi_1, \phi_2) & = (\tfrac{3\pi}{5},\tfrac{2\pi}{25})  \quad &&  V = 2\Lambda_3^4+\Lambda_1^4, \\
(\phi_1, \phi_2) & = (\tfrac{\pi}{15},\tfrac{3\pi}{25})  \quad &&  V = 2\Lambda_3^4+\Lambda_1^4+\Lambda_2^4, \\
(\phi_1, \phi_2) & = (\tfrac{3\pi}{5},\tfrac{7\pi}{25})  \quad &&  V = 2\Lambda_3^4+\Lambda_1^4+\Lambda_2^4, 
\end{align}
\end{subequations}
where the second to fourth solutions are saddle points if $32 \Lambda_1^4 \Lambda_3^4 - \tfrac{81}{2} \Lambda_2^4 \Lambda_3^4 - \tfrac{25}{4} \Lambda_1^4\Lambda_2^4 < 0$.
\end{itemize}
We see that elementary number theory allows us to find a few non-trivial critical points within the fundamental domain of the scalar potential. This search `algorithm' can be easily generalized to a theory living in a higher dimensional field space and with an arbitrary sum of cosines and sines. With the knowledge of the obtained set of critical points within the fundamental domain, one has a good starting grid that might speed up the numerical search of e.g. other saddle points or local minima where inflation can start or end. The structure and method discussed here for finding a subset of critical points may also prove useful in more general multi-axion settings as e.g.~\cite{Bachlechner:2019vcb}.

	\bibliography{RefsHybrid.bib}

	\bibliographystyle{jhep}
	
\end{document}